\documentclass[12pt,preprint]{aastex}

\shorttitle{Stellar Evolution Models for the ACS GGC Survey}
\shortauthors{Dotter et al.}

\newcommand{\Ms}{\mathrm{M_{\odot}}}
\newcommand{\Ls}{\mathrm{L_{\odot}}}
\newcommand{\Rs}{\mathrm{R_{\odot}}}
\newcommand{\Teff}{\mathrm{T_{eff}}}
\newcommand{\feh}{\mathrm{[Fe/H]}}
\newcommand{\afe}{\mathrm{[\alpha/Fe]}}

\begin{document}

\title{An ACS Survey of Galactic Globular Clusters. II.
\thanks{
Based on observations with the NASA/ESA {\it Hubble Space Telescope},
obtained at the Space Telescope Science Institute, which is operated
by AURA, Inc., under NASA contract NAS 5-26555, under programs
GO-10775 (PI: Sarajedini).
}
\\ Stellar Evolution Tracks, Isochrones, Luminosity Functions, and Synthetic Horizontal Branch Models}
\author{Aaron Dotter and Brian Chaboyer}
\affil{Department of Physics and Astronomy, Dartmouth College, 6127 Wilder Laboratory,
 Hanover, NH 03755}

\author{Darko Jevremovi\'c\altaffilmark{1} and E. Baron\altaffilmark{2}}
\affil{Homer L. Dodge Department of Physics and Astronomy, University of Oklahoma, 440 West Brooks, Room 100, Norman, OK 73019-2061}
\altaffiltext{1}{Current Address: Astronomical Observatory, Volgina 7, 11160 Belgrade, Serbia}
\altaffiltext{2}{Computational Research Division, Lawrence Berkeley National Laboratory, MS 50F-1650, 1 Cyclotron Rd, Berkeley, CA 94720}

\author{Jason W. Ferguson}
\affil{Physics Department, Wichita State University, Wichita, KS 67260-0032}

\author{Ata Sarajedini} 
\affil{Department of Astronomy, University of Florida, 211 Bryant Space Science Center, Gainesville, FL 32611}

\author{Jay Anderson}
\affil{Department of Physics and Astronomy, Rice University MS-108, Houston, TX 77005 }

\begin{abstract}
The ACS Survey of Galactic Globular Clusters, an HST Treasury Project, will deliver high quality, homogeneous photometry of 65 globular clusters.  This paper introduces a new collection of stellar evolution tracks and isochrones suitable for analyzing the ACS Survey data. Stellar evolution models were computed at $\feh$= -2.5, -2.0, -1.5, -1.0, -0.5, and 0; $\afe$= -0.2, 0, 0.2, 0.4, 0.6, and 0.8; and three initial He abundances for masses from 0.1 to 1.8 $\Ms$ and ages from 2 to 15 Gyr.  Each isochrone spans a wide range in luminosity from $M_V\sim$14 up to the tip of the red giant branch.  These are complemented by a set of He-burning tracks that extend from the zero age horizontal branch to the onset of thermal pulsations on the asymptotic giant branch.  In addition, a set of computer programs are provided that make it possible to interpolate the isochrones in $\feh$, generate luminosity functions from the isochrones, and create synthetic horizontal branch models.  The tracks and isochrones have been converted to the observational plane with two different color-$\Teff$ transformations, one synthetic and one semi-empirical, in ground-based B, V, and I, and F606W and F814W for both ACS-WFC and WFPC2 systems. All models and programs presented in this paper are available from \url{http://stellar.dartmouth.edu/$\sim$models/} and the Multimission Archive at the Space Telescope Science Institute (MAST; \url{http://archive.stsci.edu}).
\end{abstract}


\section{Introduction}
The ACS Survey of Galactic Globular Clusters \citep{sar}, (Anderson et al. 2007, in preparation) will set photometric standards for years to come.  The survey has obtained high quality photometry of 65 globular clusters from the horizontal branch down the main sequence to $\sim$0.2 $\Ms$ at the least.  The data from this survey will allow for revised absolute and differential age analyses, studies of the stellar luminosity and mass functions and dynamical evolution of a large number of clusters in a homogeneous data set among many other things. The data will provide a critical test for some aspects of current stellar evolution calculations, in particular the shape and luminosity function of the lower main sequence.  The survey will eventually combine ACS-WFC and WFPC2 photometry in the F606W and F814W bands and ground-based B, V, and I photometry for most of the clusters. In order to analyze these data, it is preferable to have theoretical models transformed to the appropriate photometric system that can be applied directly to the observations.  \citet{sir} provide a detailed analysis of all aspects of the ACS, including filter transmission curves that account for the performance of the system from optics to electronics, thereby allowing accurate synthetic photometry to be produced from synthetic fluxes \citep[e.g.]{bed}.

There are a number of groups currently producing isochrones similar to those introduced in this paper, chief among these are \citet[hereafter Victoria-Regina]{ber,vdb2}, \citet[hereafter Padova]{gir,bon,gir2}, \citet[hereafter BaSTI]{piet,piet2} and \citet{cord}, and \citet[hereafter Y$^2$ (Yonsei-Yale)]{yi2,yi,dem}. Among these, all provide isochrones for ground based B, V, and I bands but only the Padova and BaSTI libraries include isochrones on the HST photometric systems. Of these, only the Padova library extends down past the lower mass limit of the ACS survey data; the other terminate at 0.4 $\Ms$ (Y$^2$) or 0.5 $\Ms$ (Victoria-Regina and BaSTI). 

In order to analyze the full complement of photometry a new set of stellar evolution models have been produced.  This paper describes in detail the stellar evolution code used to create the models in $\S$2; the color-$\Teff$ transformations used to convert the models to the observational plane in $\S$3; properties of the model database, including descriptions of the tracks, isochrones, luminosity functions, and synthetic horizontal branch models, in $\S$4; comparisons of the isochrones with other sets isochrones and data from the ACS survey in $\S$5; and finally the results are summarized in $\S$6. Following the main text is an appendix that describes in detail the stellar evolution track, isochrone, luminosity function, and synthetic horizontal branch model files and outlines the use of computer codes distributed with the models.

\section{The Dartmouth Stellar Evolution Program}
The stellar evolution code described here is the Dartmouth Stellar Evolution Program (DSEP). Many details of DSEP have been previously described by \citet{cha,bjork}. This section is intended to explain the physics and numerical methods used in the code that have not been covered by \citet{cha,bjork}.

Stellar evolution models have been created for $\feh$=-2.5, -2.0, -1.5, -1.0, -0.5, and 0 with $\afe$=-0.2, 0, 0.2, 0.4, 0.6, and 0.8.  In the models, $\alpha$-enhancement refers to enhancements in the following $\alpha$-capture elements: O, Ne, Mg, Si, S, Ca, and Ti by the same amount as specified in $\afe$. The models assume the initial He mass fraction follows Y=0.245+1.54~Z where Y=0.245 is the primordial He abundance from Big Bang nucleosynthesis \citep{sper} and the required initial He abundance of the calibrated solar model is Y=0.274. In addition, sets of models were produced with Y$_{init}$=0.33 and 0.4 for the same range of $\feh$ with $\afe$=0 and 0.4. Complete details of the initial compositions for all models are listed in Table \ref{tab2}.

For future reference, the solar abundances were adopted from \citet{gre} and global solar parameters luminosity, radius, surface composition, and age from \citet{bah}.  Table \ref{tab1} defines the requirements for a calibrated solar model and gives the values obtained from the DSEP solar-calibrated model. The calibrated solar model has a mixing length parameter $\alpha_{ML}$=1.938 with initial composition X=0.7071 and Z=0.01885. Further details of the physics employed by DSEP can be found in the remainder of this section and the two papers referenced in the first paragraph of this section paragraph.

\clearpage

%TABLE 1
\begin{deluxetable}{ll}
\tablecolumns{2}
\tablewidth{0pc}
\tablecaption{Solar Parameters\label{tab1}}
\tablehead{\colhead{Adopted Value}&\colhead{Solar Model Value}}
\startdata
Age = 4.57 Gyr & \nodata \\
$\Ms$ = 1.9891$\times$10$^{33}$ g & \nodata  \\
$\Rs$ = 6.9598$\times$10$^{10}$ cm & Log(R/$\Rs$)=0.0001  \\ 
$\Ls$ = 3.8418$\times$10$^{33}$ erg/s & Log(L/$\Ls$)=0.0005  \\
R$_{CZ}$/$\Rs$ = 0.713 $\pm$ 0.001 & 0.716  \\
Z/X = 0.0229 & 0.02288
\enddata
\tablerefs{\citet{bah,gre}}
\end{deluxetable}

%TABLE 2
\begin{deluxetable}{cccccccccccc}
\rotate
%\tabletypesize{\footnotesize}
\tablecolumns{12}
\tablewidth{0pc}
\tablecaption{Initial Compositions\label{tab2}}
\tablehead{
\colhead{}&\colhead{}&\multicolumn{6}{c}{Y=0.245+1.54Z}&\multicolumn{2}{c}{Y=0.33}&\multicolumn{2}{c}{Y=0.4}\\
\cline{3-12}\\
\colhead{$\feh$}&\colhead{$\afe$=}&\colhead{-0.2}&\colhead{0}&\colhead{0.2}&\colhead{0.4}&\colhead{0.6}&\colhead{0.8}&\colhead{0}&\colhead{0.4}&\colhead{0}&\colhead{0.4}
}
\startdata
-2.5 & X= & 0.7549 & 0.7548 & 0.7548 & 0.7547 & 0.7546 & 0.7543 & 0.6699 & 0.6699 & 0.6000 & 0.5999 \\
\phn & Z= & 4.09E-5 & 5.48E-5 & 7.63E-5 & 1.11E-4 & 1.65E-4 & 2.52E-4 & 4.85E-5 & 9.85E-5 & 4.34E-5 & 8.82E-5 \\
-2.0 & X= & 0.7547 & 0.7545 & 0.7544 & 0.7541 & 0.7536 & 0.7529 & 0.6698 & 0.6697 & 0.5999 & 0.5997 \\
\phn & Z= & 1.30E-4 & 1.72E-4 & 2.41E-4 & 3.50E-4 & 5.24E-4 & 7.97E-4 & 1.53E-4 & 3.50E-4 & 1.37E-4 & 2.79E-4 \\
-1.5 & X= & 0.7539 & 0.7536 & 0.7530 & 0.7521 & 0.7507 & 0.7484 & 0.6695 & 0.6690 & 0.5996 & 0.5991 \\
\phn & Z= & 4.10E-4 & 5.47E-4 & 7.62E-4 & 1.11E-3 & 1.65E-3 & 2.50E-3 & 4.85E-4 & 9.83E-4 & 4.34E-4 & 8.81E-4 \\
-1.0 & X= & 0.7516 & 0.7505 & 0.7487 & 0.7459 & 0.7415 & 0.7346 & 0.6685 & 0.6669 & 0.5986 & 0.5972 \\
\phn & Z= & 1.29E-3 & 1.72E-3 & 2.40E-3 & 3.46E-3 & 5.15E-3 & 7.77E-3 & 1.53E-3 & 3.46E-3 & 1.37E-3 & 2.78E-3 \\
-0.5 & X= & 0.7444 & 0.7441 & 0.7355 & 0.7270 & 0.7139 & 0.6942 & 0.6652 & 0.6603 & 0.5957 & 0.5913 \\
\phn & Z= & 4.05E-3 & 5.37E-3 & 7.44E-3 & 1.07E-2 & 1.57E-2 & 2.32E-2 & 4.82E-3 & 9.70E-3 & 4.31E-3 & 8.69E-3 \\
 0.0 & X= & 0.7174 & 0.7071 & 0.6882 & 0.6628 & 0.6247 & 0.5710 & 0.6550 & 0.6402 & 0.5866 & 0.5734 \\
\phn & Z= & 1.43E-2 & 1.89E-2 & 2.55E-2 & 3.52E-2 & 4.97E-2 & 7.02E-2 & 1.50E-2 & 2.98E-2 & 1.34E-2 & 2.66E-2
\enddata
\end{deluxetable}

\subsection{Convective Core Overshoot}

%TABLE 3
\begin{deluxetable}{cccccccccccc}
%\rotate
\tabletypesize{\footnotesize}
\tablecolumns{12}
\tablewidth{0pc}
\tablecaption{Minimum Mass in $\Ms$ for Convective Core Overshoot\label{tab3}}
\tablehead{
\colhead{}&\colhead{}&\multicolumn{6}{c}{Y=0.245+1.54Z}&\multicolumn{2}{c}{Y=0.33}&\multicolumn{2}{c}{Y=0.4}\\
\cline{3-12}\\
\colhead{$\feh$}&\colhead{$\afe$=}&\colhead{-0.2}&\colhead{0}&\colhead{0.2}&\colhead{0.4}&\colhead{0.6}&\colhead{0.8}&\colhead{0}&\colhead{0.4}&\colhead{0}&\colhead{0.4}
}
\startdata
-2.5 & \phn & 1.6 & 1.6 & 1.6 & 1.5 & 1.3 & 1.3 & 1.5 & 1.5 & 1.5 & 1.5 \\ 
-2.0 & \phn & 1.4 & 1.4 & 1.4 & 1.4 & 1.3 & 1.3 & 1.5 & 1.5 & 1.5 & 1.5 \\
-1.5 & \phn & 1.4 & 1.4 & 1.4 & 1.4 & 1.3 & 1.3 & 1.4 & 1.4 & 1.4 & 1.4 \\
-1.0 & \phn & 1.4 & 1.4 & 1.4 & 1.4 & 1.2 & 1.2 & 1.4 & 1.4 & 1.4 & 1.4 \\
-0.5 & \phn & 1.2 & 1.2 & 1.2 & 1.2 & 1.2 & 1.0 & 1.2 & 1.2 & 1.2 & 1.2 \\
 0.0 & \phn & 1.2 & 1.2 & 1.1 & 1.1 & 1.0 & 0.9 & 1.2 & 1.1 & 1.2 & 1.1 \\
\enddata
\end{deluxetable}

\clearpage

The adopted convective core overshoot (CCO) treatment is that of \citet{dem}.  The basic idea behind the method is that the extent of CCO depends on the size of the convective core.  The lowest mass main sequence stars that develop convective cores have relatively small ones.  As stellar mass increases beyond this minimum, the convective cores become fully established and the amount of overshoot increases.

The extent of CCO is treated as a fraction of the pressure scale height at the core boundary. At the minimum mass (listed for all compositions in Table \ref{tab3}) the amount of CCO is 5\% of the pressure scale height.  Between the minimum and 0.2 $\Ms$ above the minimum, the extent of CCO is ramped linearly from 5\% to 20\%. For masses greater than 0.2 $\Ms$ above the minimum, the extent of CCO is held constant at 20\% of the pressure scale height. CCO plays a role in only the youngest isochrones presented in this paper, those with ages less than about 5 Gyr.

\subsection{Microscopic Diffusion and Gravitational Settling}
\citet{cha} described the implementation of diffusion and gravitational settling in DSEP; the following is a brief summary.  The basic formalism is that of \citet{tho} but diffusion has been inhibited in the outer 0.1 $\Ms$. In the outermost 0.05 $\Ms$ diffusion is completely stopped and in the 0.05 $\Ms$ below that diffusion is linearly ramped from zero to the full effect. This approach is meant to satisfy the lack of observational evidence for diffusion in metal poor stars \citep[e.g.]{grat,ram}, though recently evidence for some diffusion has been reported in NGC 6397 \citep{korn}.

\subsection{Equation of State}
The basic equation of state (EOS) employed by DSEP is a general ideal gas equation of state with the Debye-H\"uckel correction and performs sufficiently well for stellar models with masses greater than about 0.7 $\Ms$ \citep{cha2}. Because this EOS neglects many non-ideal effects that become important at lower temperatures and higher densities (such as are found in low mass main sequence stars) the basic EOS was used only for models with masses greater than 0.8 $\Ms$.

For stellar models with 0.1 $\leq$ M $\leq$ 0.8 $\Ms$ the detailed EOS code developed by \citet{irwin}, FreeEOS (in the EOS4 configuration), was used.  FreeEOS takes into account many non-ideal effects and is generally superior for high density, low temperature environments.  Furthermore, FreeEOS takes into account the makeup of Z. Stellar models produced with this EOS are in good agreement with observations of low mass stars (see $\S$5).  Irwin has written extensively about the numerical methods and physical models employed by FreeEOS; the interested reader should consult the FreeEOS website\footnote{http://freeeos.sourceforge.net}.

Beginning at approximately 0.7 $\Ms$, stellar evolution tracks produced with FreeEOS and the default EOS begin to differ in the H-R diagram.  As stellar mass decreases, these differences increase, becoming substantial at 0.1 $\Ms$.  The current version FreeEOS (2.0) does not operate in the parameter range occupied by main sequence models below 0.1 $\Ms$.

\subsection{Opacities}
Opacities were adopted from two sources\footnote{OPAL opacities---http://www-phys.llnl.gov/Research/OPAL/\\\citet{fer} opacities---http://webs.wichita.edu/physics/opacity/}.  In both cases, separate sets of tables were generated for each of the $\afe$ mixtures listed above.  The OPAL opacities \citep{igl} were used at high temperature (log T $>$ 4.5) while opacities computed for this effort following \citet{fer} were used at low temperature (log T $<$ 4.3).  Between log T = 4.3 and 4.5 the opacity was ramped between the two sources.  \citet{fer} have shown that OPAL opacities agree with their own to within 5\% in this temperature range.

During core and shell He fusion (HB and AGB evolution) the OPAL Type 2 opacity tables were used.  The Type 2 opacity tables allow mass fractions of two individual elements (in this case C and O) to vary independent of overall Z.  These opacity tables are particularly helpful in the case of He-burning evolution where the He core gradually becomes a mixture of almost entirely C and O over the course of $\sim$100 Myr.

Conductive opacities were adopted from \citet{hubb} in the non-relativistic limit and \citet{canu} in the relativistic limit via the fitting formulas of \citet{swei}.

\subsection{Surface Boundary Conditions}
It is typical in stellar evolution codes to see surface boundary conditions treated as simple mathematical models of surface temperature as a function of optical depth.  The almost universal choices are the Eddington approximation or the empirical fit to the Sun done by \citet{kri}.

However, particularly in the case of cool, low mass main sequence stars it is common practice to derive surface boundary conditions from model atmosphere calculations \citep{bar}. Model atmosphere boundary conditions are an important ingredient in models of cool or very low mass stars (M $<$ 0.4 $\Ms$) but make little or no difference for hotter, more massive stars.

DSEP has been modified to accept surface boundary conditions derived from PHOENIX model atmospheres \citep{phxa,phxb}.  One of the strengths of the PHOENIX code is attention to the effects of molecules and grains that appear at low temperature. Indeed, the low temperature opacities used in this paper were computed with a modified version of the PHOENIX code \citep{fer}. A grid of stellar atmosphere models was produced covering the range of $\feh$ and $\afe$ listed above based on the \citet{gre} solar abundances.  The grid spans surface gravity -0.5 $\leq$ log g $\leq$ 5.5 and effective temperature 2,500 K $\leq$ $\Teff$ $\leq$ 10,000 K.  For the hottest tracks, model atmosphere grids from \citet{cas} were supplemented for $\Teff$ $>$ 10,000 K.  A coarser grid of model atmospheres was generated to explore the effect of enhancing the He abundance on atmosphere structure and synthetic flux.  While additional He in the atmosphere has a measurable effect on the gas pressure there is essentially no change in the synthetic flux.  Because He is a noble gas, it is difficult to excite for $\Teff$ below 10,000 K and thus contributes little to the flux.  At present, He-enhancement is treated as an adjustment to the gas pressure that serves as the surface boundary condition but is ignored in the synthetic colors.

The surface boundary condition is satisfied by specifying pressure as a function of $\Teff$ and log g. To obtain a pressure from each model atmosphere, the location within the atmosphere structure where T=$\Teff$ was found and the gas pressure extracted from that location. This approach differs from those in which the pressure is extracted from the model atmosphere at constant optical depth. Before deciding to follow the adopted method, tests were performed by comparing surface boundary condtions extracted from the atmospheres at constant optical depth ($\tau$=2/3 and 10) but the tests showed that the T=$\Teff$ method produced the best results when compared with observations. The physical approximations made by model atmosphere and stellar evolution codes are different enough than any choice of where to connect the interior to the atmosphere will encounter some inconsistencies.

%FIGURE 1
\begin{figure}
\epsscale{1.0}
\plotone{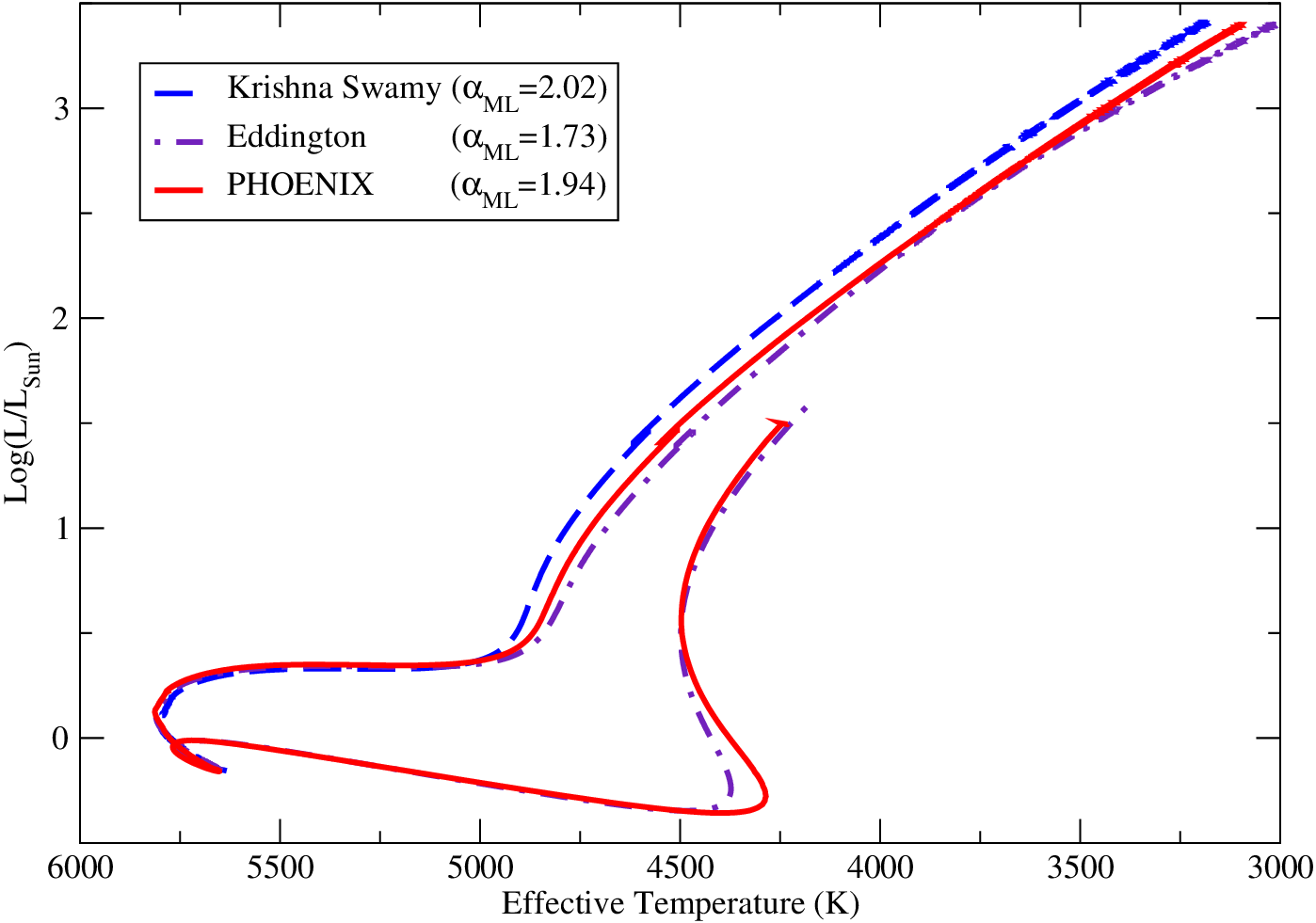}
\caption{The evolution of calibrated solar models from the pre-main sequence to the tip of the red giant branch the in the theoretical H-R diagram (except for K-S where no pre-main sequence is plotted). Each model employs a different surface boundary condition and requires a different solar-calibrated mixing length. The PHOENIX model atmosphere boundary condition adopted in this paper lies between the Eddington and Krishna Swamy models both in terms of the mixing length and, as a result, the temperature of the red giant branch.  Note also that while the Eddington and Krishna Swamy models have a similar amount of curvature above log L/$\Ls$=1 the PHOENIX model red giant branch is relatively straight.}
\label{sbc}
\end{figure}

Calibrated solar models using the Eddington, Krishna Swamy, and PHOENIX surface boundary conditions were produced in order to compare the three boundary conditions in the H-R diagram.  The Krishna Swamy model required the largest mixing length ($\alpha_{ML}$=2.02) and has the hottest red giant branch.  The Eddington model required the smallest mixing length ($\alpha_{ML}$=1.73) and has the coolest red giant branch.  The PHOENIX model was located in between the others ($\alpha_{ML}$=1.94) with a slightly flatter red giant branch in the H-R diagram (see Figure \ref{sbc}).

\subsection{Nuclear Reactions and Neutrino Cooling Rates}
Nuclear reaction rates were adopted from \citep{adel} with two exceptions: first, the $^{14}$N(p,$\gamma$)$^{15}$O reaction \citep{imbr} and second, the $^{12}$C($\alpha$,$\gamma$)$^{16}$O reaction \citep{kunz}. Neutrino cooling rates were adopted from \citet{haft}.

\subsection{He-Burning Evolution}

%TABLE 4
\begin{deluxetable}{cccccccccccc}
\rotate
%\tabletypesize{\footnotesize}
\tablecolumns{12}
\tablewidth{0pc}
\tablecaption{Surface Composition for Zero Age Horizontal Branch Models\label{tab4}}
\tablehead{
\colhead{}&\colhead{}&\multicolumn{6}{c}{Y$_{init}$=0.245+1.54Z$_{init}$\tablenotemark{\dagger}}&\multicolumn{2}{c}{Y$_{init}$=0.33}&\multicolumn{2}{c}{Y$_{init}$=0.4}\\
\cline{3-12}\\
\colhead{$\feh$}&\colhead{$\afe$=}&\colhead{-0.2}&\colhead{0}&\colhead{0.2}&\colhead{0.4}&\colhead{0.6}&\colhead{0.8}&\colhead{0}&\colhead{0.4}&\colhead{0}&\colhead{0.4}
}
\startdata
%                   |                     Y=0.245+1.54*Z              |    Y=0.33         |     Y=0.40         |
%FeH   aFe=-0.2       0         0.2       0.4       0.6       0.8       0.0       0.4       0.0       0.4
-2.5 & X= & 0.7564  & 0.7565  & 0.7561  & 0.7549  & 0.7549  & 0.7546  & 0.6835  & 0.6828  & 0.5967  & 0.6212  \\  %done
\phn & Z= & 3.97E-5 & 5.31E-5 & 7.40E-4 & 1.60E-4 & 1.08E-4 & 2.44E-4 & 4.63E-5 & 9.40E-5 & 8.40E-3 & 8.33E-5 \\

-2.0 & X= & 0.7547  & 0.7548  & 0.7551  & 0.7541  & 0.7524  & 0.7510  & 0.6820  & 0.6810  & 0.6208  & 0.6193  \\  %done
\phn & Z= & 1.26E-4 & 1.67E-4 & 2.33E-4 & 3.41E-4 & 5.06E-4 & 7.69E-4 & 1.46E-4 & 2.97E-4 & 1.30E-4 & 2.64E-4 \\

-1.5 & X= & 0.7529  & 0.7517  & 0.7509  & 0.7491  & 0.7464  & 0.7427  & 0.6785  & 0.6763  & 0.6180  & 0.6156  \\  %done
\phn & Z= & 3.96E-4 & 5.29E-4 & 7.37E-4 & 1.07E-3 & 1.60E-3 & 2.43E-3 & 4.64E-4 & 9.44E-4 & 4.12E-4 & 8.37E-4 \\

-1.0 & X= & 0.7472  & 0.7453  & 0.7423  & 0.7383  & 0.7325  & 0.7238  & 0.6732  & 0.6698  & 0.6125  & 0.6076  \\  %done
\phn & Z= & 1.25E-3 & 1.67E-3 & 2.33E-3 & 3.37E-3 & 5.02E-3 & 7.59E-3 & 1.47E-3 & 3.00E-3 & 1.31E-3 & 2.67E-3 \\  

-0.5 & X= & 0.7350  & 0.7309  & 0.7242  & 0.71485 & 0.7010  & 0.6810  & 0.6641  & 0.6600  & 0.6034  & 0.5967  \\  %done
\phn & Z= & 3.94E-3 & 5.24E-3 & 7.27E-3 & 1.05E-2 & 1.54E-3 & 2.28E-2 & 4.67E-3 & 9.48E-3 & 4.16E-3 & 8.41E-3 \\  
 
 0.0 & X= & 0.7166  & 0.6928  & 0.6741  & 0.6497  & 0.6131  & 0.5640  & 0.6477  & 0.6303  & 0.5877  & 0.5854  \\  %done
\phn & Z= & 1.41E-2 & 1.85E-2 & 2.50E-2 & 3.46E-2 & 4.88E-2 & 6.87E-2 & 1.47E-2 & 2.91E-2 & 1.31E-2 & 2.68E-2     
\enddata
\tablenotetext{\dagger}{Y$_{init}$ and Z$_{init}$ refer to the initial Y and Z values, for the initial compositions see Table \ref{tab2}. The Z values in this table differ from those in Table \ref{tab2} as a result of diffusion.}
\end{deluxetable}

%TABLE 5
\begin{deluxetable}{cccccccccccc}
%\rotate
\tabletypesize{\footnotesize}
\tablecolumns{12}
\tablewidth{0pc}
\tablecaption{He Core Mass in $\Ms$ for Zero Age Horizontal Branch Models\label{tab5}}
\tablehead{
\colhead{}&\colhead{}&\multicolumn{6}{c}{Y=0.245+1.54Z}&\multicolumn{2}{c}{Y=0.33}&\multicolumn{2}{c}{Y=0.4}\\
\cline{3-12}\\
\colhead{$\feh$}&\colhead{$\afe$=}&\colhead{-0.2}&\colhead{0}&\colhead{0.2}&\colhead{0.4}&\colhead{0.6}&\colhead{0.8}&\colhead{0}&\colhead{0.4}&\colhead{0}&\colhead{0.4}
}
\startdata
-2.5 & \phn & 0.506 & 0.505 & 0.504 & 0.502 & 0.500 & 0.497 & 0.489 & 0.485 & 0.474 & 0.473 \\
-2.0 & \phn & 0.500 & 0.499 & 0.498 & 0.494 & 0.492 & 0.491 & 0.483 & 0.479 & 0.469 & 0.468 \\
-1.5 & \phn & 0.495 & 0.493 & 0.491 & 0.489 & 0.488 & 0.486 & 0.478 & 0.474 & 0.464 & 0.463 \\
-1.0 & \phn & 0.489 & 0.488 & 0.486 & 0.484 & 0.481 & 0.478 & 0.473 & 0.469 & 0.460 & 0.458 \\
-0.5 & \phn & 0.483 & 0.482 & 0.480 & 0.477 & 0.473 & 0.469 & 0.469 & 0.465 & 0.457 & 0.455 \\
 0.0 & \phn & 0.475 & 0.473 & 0.469 & 0.464 & 0.458 & 0.447 & 0.465 & 0.460 & 0.454 & 0.451
\enddata
\end{deluxetable}

Stellar evolution calculations that follow the He flash in low mass stars are possible but numerically difficult and time consuming.  Furthermore, studies have shown that so-called pseudo-ZAHB (zero age horizontal branch) models that are concocted from red giant branch tip models are nearly indistinguishable from full He flash calculations in terms of the subsequent evolution \citep{vdb3,pier,ser}.  Beginning with a number of these pseudo-ZAHB models a complete set of core and shell He-burning tracks were created based on the red giant branch tip surface compositions listed in Table \ref{tab4} and He core masses listed in Table \ref{tab5}.  Since He core mass varies slightly with age or stellar mass, the core mass was taken from the stellar track that reached the red giant branch tip closest to 12 Gyr in each case, if no track for a given composition reached the tip within 11-13 Gyr the surface composition and He core mass were interpolated between the two nearest tracks. The core and shell He-burning tracks allow for the population of the horizontal branch and asymptotic giant branch through the use of individual tracks or the construction of synthetic horizontal branch models (for more on this see $\S$4.4). The remainder of this section describes the physics in DSEP specific to core and shell He-burning processes.

Semiconvection has long been known to play an important role in stars with steep composition gradients such as occur in stars undergoing core He fusion. Instead of including a full treatment of a semiconvection zone, DSEP uses a simple model to treat the expansion of the core boundary during He-burning developed by \citet{cast}. Thermal energies have been artificially reduced when the core He mass fraction falls below 0.1 to discourage small loops in the H-R diagram known as breathing pulses as suggested by A. Sweigart (private communication) and first noted by \citet{dorm}.  Breathing pulses occur when the convective core expands outward into He-rich material and are widely believed to be unphysical artifacts of the models. Neglecting thermal energies in the core during the final 10\% of core He fusion effectively freezes the size of the core thereby removing breathing pulses.

High temperature opacities that allow for varying levels of C and O (as described in $\S$2.4) allow for correct opacity coverage of stellar cores during the entire core and shell He fusion phases.

Time steps during core He fusion are based on the depletion of He in the core such that the mass fraction Y$_{core}$ cannot fall by more than a fixed amount during each time step.  When Y$_{core}$ falls to 0.1 then the time step is set by the percentage change in Y$_{core}$ rather than by a a fixed amount.  During shell He fusion (the AGB phase) time steps are further constrained by the total He-burning luminosity: the sum of the triple-$\alpha$ and $^{12}$C($\alpha$,$\gamma$)$^{16}$O luminosities.  These time steps are used in addition to the time steps that govern the H-shell.

\section{Color-$\Teff$ Transformations}
The stellar evolution models described in the last section have been transformed to the observational plane by two different methods.  One is based on the semi-empirical colors and bolometric corrections of \citet{vdb} and the transformation equations found in Appendix D of \citet{sir}, specifically those in Tables 22 and 24, hereafter referred to as the Empirical transformation.  The other is based entirely on synthetic fluxes from PHOENIX (the same model atmosphere grids described in $\S$2.5) with the appropriate filter transmission curves and is hereafter referred to as the Synthetic transformation.

Both transformations have been normalized to Vega and include ground-based B, V and I magnitudes on the \citet{bes} system and the HST filters F606W and F814W on both the WFPC2 and ACS-WFC systems from \citet{sir}\footnote{Readers interested in obtaining isochrones in other filters should contact the authors.}.

\subsection{Empirical Transformation}
\citet{vdb} introduced a new set of semi-empirical color transformations and bolometric corrections for standard B, V, R, and I bandpasses.  The colors and bolometric corrections were empirically constrained so that the isochrones of \citet{ber} were able to accurately reproduce the color-magnitude diagrams (CMD's) of various Galactic globular, open clusters, and assorted other stars spanning the range of metallicities commonly observed in the Milky Way.  The BC$_V$, B-V, and V-I of \citet{vdb} make up the ground-based portion of the Empirical transformation.

As part of an exhaustive study of the ACS, \citet{sir} derived equations to convert between ground-based BVRI magnitudes and HST ACS-WFC (and HRC) filters as well as between ACS and WFPC2 magnitudes.  The V and I magnitudes from \cite{vdb} were used to obtain ACS F606W and F814W magnitudes using the (V,V-I)$\rightarrow$F606W and (I,V-I)$\rightarrow$F814W equations.  The ACS magnitudes were then used to obtain the corresponding WFPC2 magnitudes.  The VegaMAG zeropoints were used in each case.

\subsection{Synthetic Transformation}
Filter transmission curves in B, V, and I from \citet{bes} were used along with curves from both ACS-WFC and WFPC2 filters F606W and F814W from \citet{sir} to create a synthetic color-$\Teff$ transformation.  The synthetic colors were generated with fluxes from the same PHOENIX model atmospheres used for the surface boundary conditions described in $\S$2.5.  These colors account for the effects of $\alpha$-enhancement---something not possible with the empirical transformation.  The effects are small but noticeable, particularly below $\sim$4,000 K.

As mentioned in $\S$2.5, model atmospheres with enhanced He were produced to properly treat He-enhancement in the surface boundary condition. Tests have been performed to address the possible changes to synthetic colors due to the enhancement of He in the atmosphere for the Y=0.33 and 0.40 isochrones.  As discussed in $\S$2.5, the flux is virtually unchanged by enhancing He and this carries through to the colors.  Therefore He-enhancement has been omitted from the Synthetic color transformation.

The procedure followed to derive the Synthetic transformation closely follows that of \citet{bed}.  The bolometric correction in V and all colors were normalized by assuming Vega has V=0.026 and all colors equal to zero.

\subsection{Discussion}

%FIGURE 2
\begin{figure}
\plotone{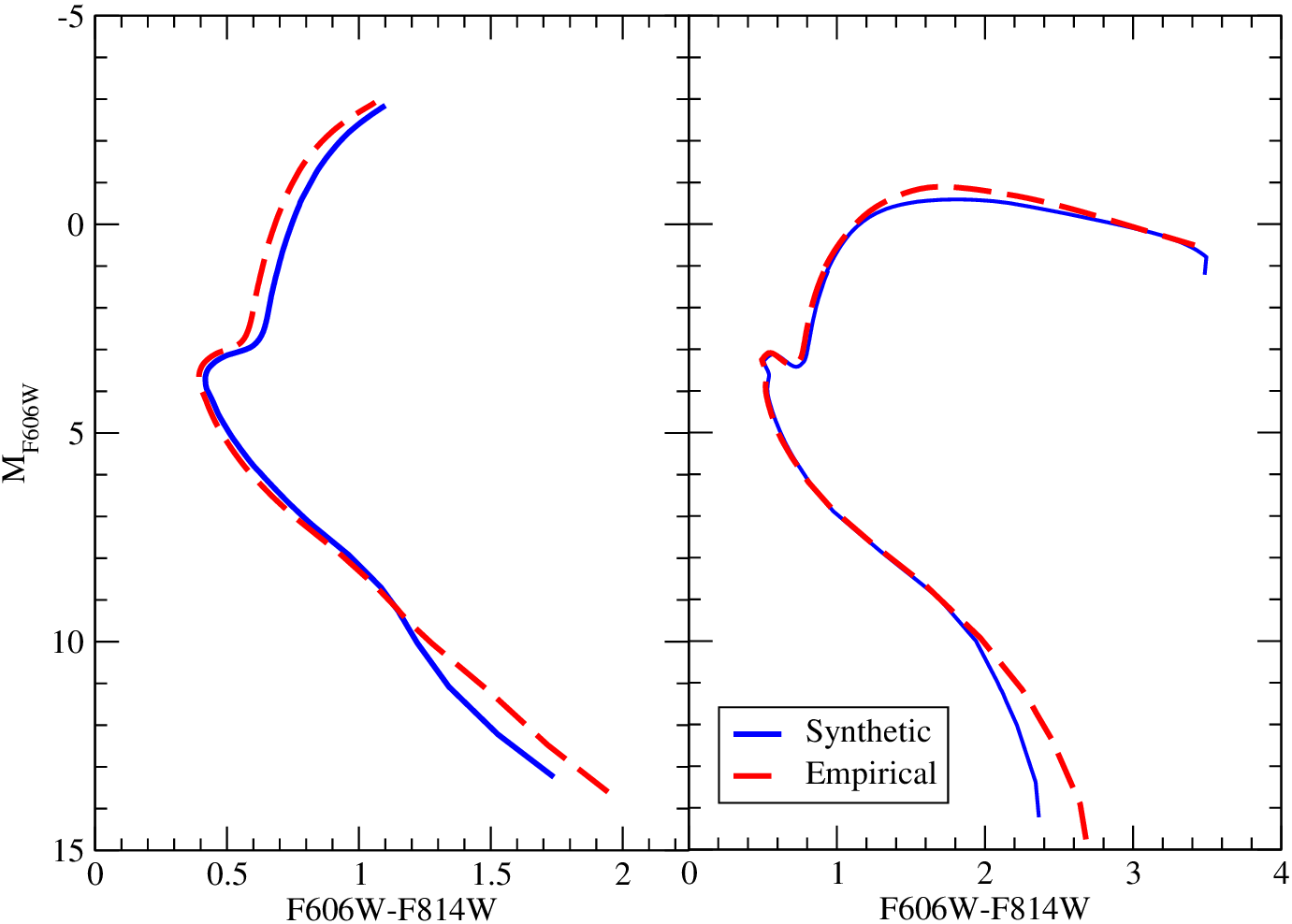}
\caption{Color comparison for the ACS bands: Left panel---Isochrones with $\feh$=-2.0, $\afe$=+0.4 at 12 Gyr.  Right panel---Isochrones with $\feh$=0, $\afe$=+0.2 at 5 Gyr. The general trend is for the Empirical colors to be bluer on the giant branch and redder on the lower main sequence than the Synthetic colors. (Empirical---dashed line; Synthetic---solid line)
\label{clr1}}
\end{figure} 

Isochrones constructed from the two color transformations tend to agree within $\sim$0.05 mag in all filters (the only noteworthy exception occurs at high Z where the Synthetic B is flawed). Figure \ref{clr1} compares the two color transformations over the full range of the isochrones for the case of $\feh$=-2.0, $\afe$=+0.4 at 12 Gyr in the left panel and $\feh$=0, $\afe$=0.2 at 5 Gyr in the right panel.  In each panel, the transformed isochrones were plotted with no adjustments in either direction, thus any difference is due entirely to the color transformation.  The Empirical isochrones (dashed line) generally have a narrower color difference between the turn off and the base of the red giant branch than the Synthetic isochrones (solid line). The red giant branches move closer together until they overlap (or nearly so) at the tip.  There is good agreement for the first $\sim$5 mag below the turn off and then the Empirical transformation moves to the red of the Synthetic.

Further side-by-side comparisons of the two color transformations for a variety of metallicities and ages, including comparisons to photometry with distance and reddening estimates, can be found in \citet{sar}.

\section{Stellar Evolution Models}
The following section lists and describes the different types of models accompanying this paper, either as data files or files that may be generated with one of several computer programs distributed along with the data files.  The contents of each type of data file and the use of the accompanying programs are described in the Appendix.

\subsection{Stellar Evolution Tracks}

%FIGURE 3
\begin{figure}
\epsscale{0.8}
\plotone{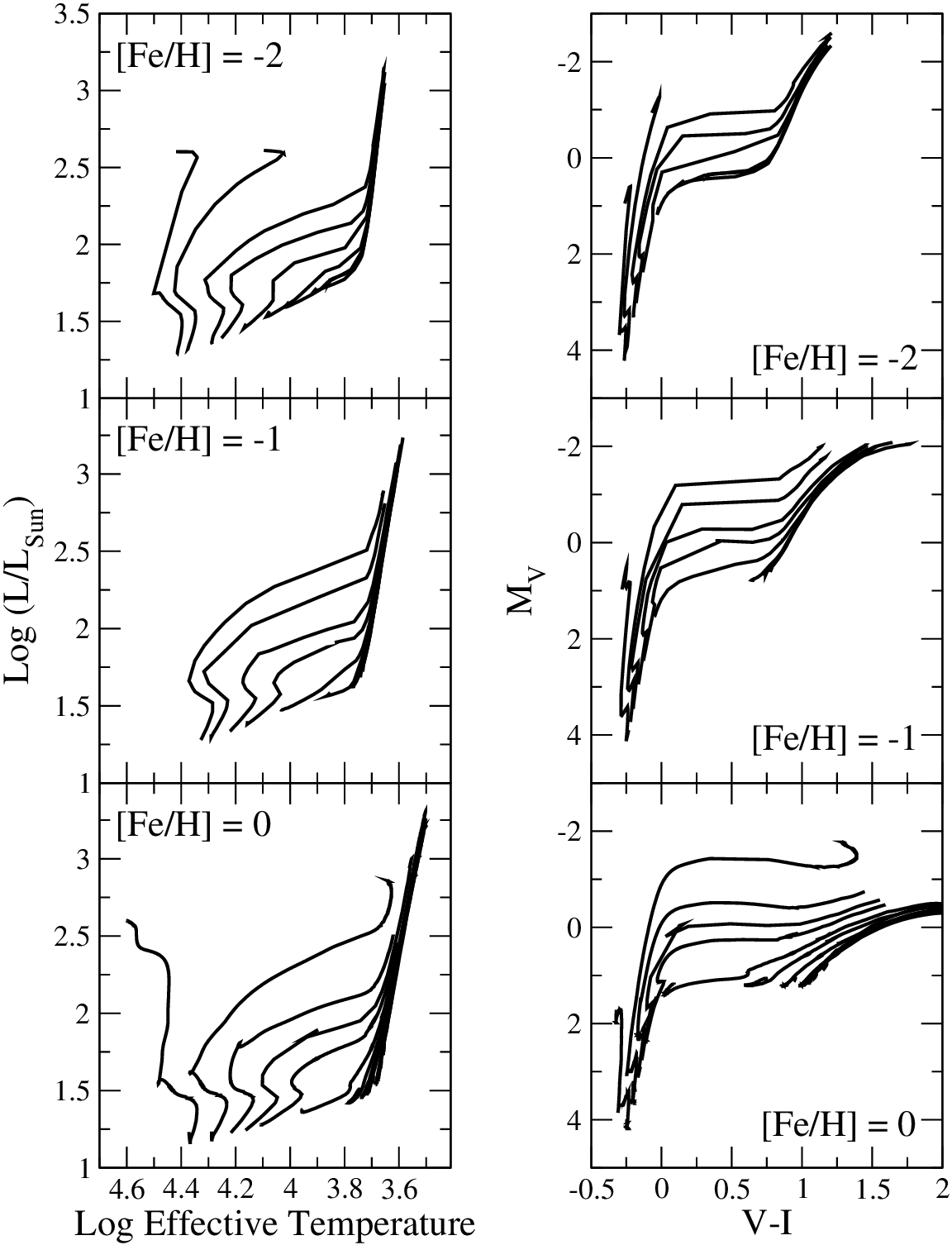}
\caption{He-burning tracks for $\feh$=-2 (top), -1 (middle), and 0 (bottom) all with scaled-solar abundances.  All masses from the minimum on the horizontal branch (see Table \ref{tab5} for He core masses) of 0.51 $\Ms$ at $\feh$=-2, 0.5 $\Ms$ at $\feh$=-1, and 0.48 $\Ms$ at $\feh$=0 up to 0.9 $\Ms$ are shown. The left side shows the Log $\Teff$ vs. Log L/$\Ls$ and the right side the V vs. V--I color-magnitude diagram using Empirical color transformation. The $\feh$=0 color-magnitude diagram has been truncated in order to show the hotter stars more clearly, the tracks extend out to V--I$\sim$4.\label{hb}}
\end{figure}

Stellar evolution tracks of individual masses for each composition are available in two formats: one listing the theoretical quantities and one listing basic global parameters and absolute magnitudes in each of the seven filters listed in the previous section.  The tracks include evolution from the pre-main sequence to the tip of the red giant branch from 0.1 $\Ms$ $<$ M $<$ 1.8 $\Ms$ in steps of 0.05 $\Ms$.  

Separate tracks follow the He-burning evolution from the ZAHB to the onset of the thermal pulsation (TP-AGB) phase; these tracks are produced at masses from $\sim$0.01 $\Ms$ above the He core mass (see Table \ref{tab5}) up to 0.51 $\Ms$ in steps of 0.01 $\Ms$, from 0.52 to 0.56 $\Ms$ in steps of 0.02 $\Ms$, from 0.6 to 0.9 $\Ms$ in steps of 0.1 $\Ms$, and finally from 1 to 1.75 $\Ms$ in steps of 0.25 $\Ms$. Figure \ref{hb} shows all of the He-burning tracks from the minimum mass up to 0.9 $\Ms$ for $\feh$=-2, -1, and 0 with scaled-solar abundances both in the theoretical H-R diagram and the V vs. V--I color-magnitude diagram.  The $\feh$=0 tracks have been truncated in the color-magnitude diagram in Figure \ref{hb} in order to more clearly show the hotter stars, these tracks extend out to V--I$\sim$4.

Each track file containing theoretical quantities lists for each time step: age in years, log effective temperature in Kelvin, log surface gravity (CGS), log luminosity in $\Ls$, log radius in $\Rs$, core He mass fraction Y$_{core}$, core heavy element mass fraction Z$_{core}$, surface heavy element to hydrogen ratio (Z/X)$_{surf}$, H luminosity in $\Ls$, He luminosity in $\Ls$, He core mass in $\Ms$, and CO core mass in $\Ms$.  All logarithms are base 10; see Table \ref{tab1} for the adopted solar values.

Each track file containing observable quantities lists at each time step the first four quantities from the theoretical track file (age, log $\Teff$, log g, and log L/$\Ls$) plus absolute magnitudes in each of the seven filters listed above. Samples of both types of track file are given in the Appendix.

\subsection{Isochrones}
Isochrones were built using the equivalent evolutionary phase (EEP) principle.   Each isochrone consists of $\sim$230 points extending from the lowest mass point at 0.1 $\Ms$ up to the tip of the red giant branch.  EEPs are assigned as a function of core He mass fraction (Y$_{core}$) on the main sequence and as a function of He core mass on the red giant branch. Isochrones are provided for ages ranging from 2 to 15 Gyr in increments of 0.5 Gyr for all compositions.

%FIGURE 4
\begin{figure}
\epsscale{0.8}
\plotone{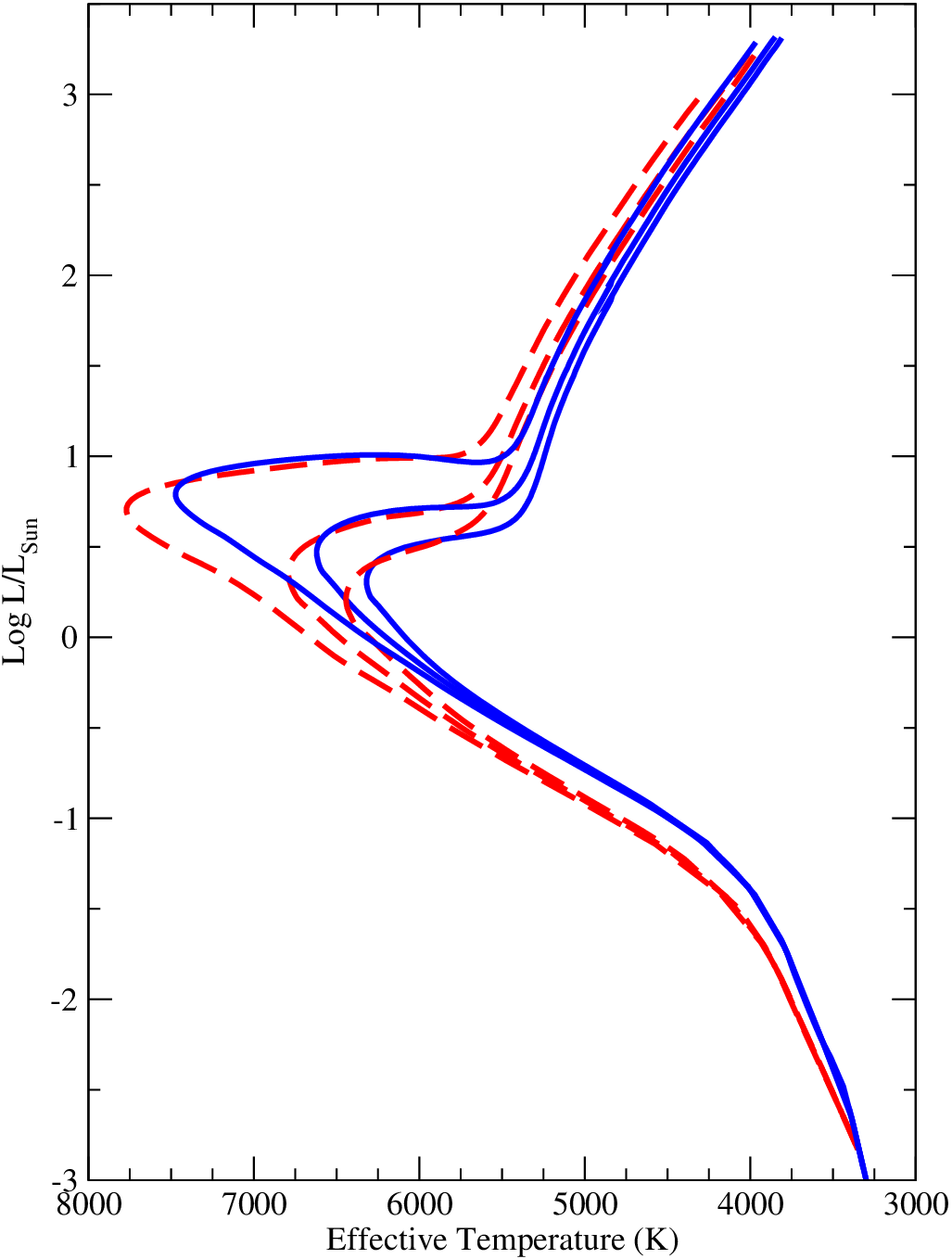}
\caption{The effect of He-enhancement at constant $\feh$=-1 and $\afe$=0 in the H-R diagram.  Ages shown are 4, 8, and 12 Gyr.  The solid lines have Y$_{init}$=0.248 while the dashed lines have Y$_{init}$=0.4.}
\label{iso_y}
\end{figure}

Figure \ref{iso_y} demonstrates the effects of He-enhancement at $\feh$=-1, $\afe$=0.  Isochrones are shown for 4, 8, and 12 Gyr with Y$_{init}\sim$0.25 (solid lines) and Y$_{init}$=0.4 (dashed lines).  Enhancing He at the expense of H decreases the opacity and increases the mean molecular weight.  The end result is an isochrone with Y=0.40 is significantly hotter at a given age and Z than its counterpart with Y$\sim$0.25 as is demonstrated in Figure \ref{iso_y}.

The model grid has a relatively fine spacing in both age and $\afe$, such that parameters of clusters being analyzed can be well constrained by the ages and $\afe$ values provided.  The only parameter that could be more tightly constrained by observations is [Fe/H] and so a computer program has been distributed with the isochrones to allow interpolation in $\feh$ (see the Appendix for details).  The interpolation code has been extensively tested and provides results that are reasonably accurate over the range of metallicities provided in the isochrone database.

\subsection{Luminosity Functions}
%FIGURE 5
\begin{figure}
\plotone{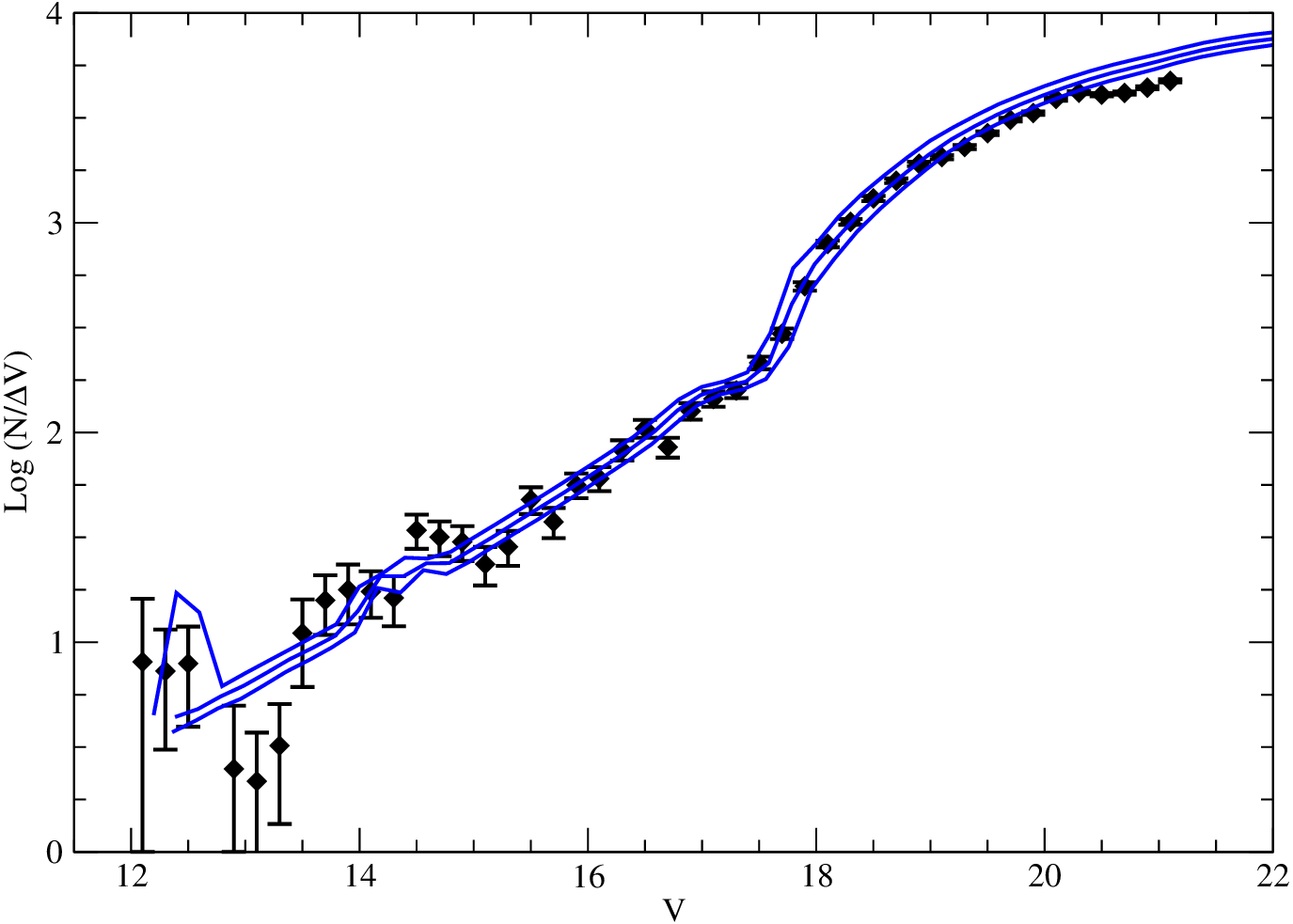}
\caption{Example of theoretical luminosity functions compared to the observed luminosity function of M 92 \citep{paust}.  The theoretical luminosity functions shown were derived from Synthetic color isochrones at $\feh$=-2.3 and $\afe$=0.4 for ages 12, 13, and 14 Gyr with a Salpeter IMF  (see also Figure \ref{m92}).  The assumed distance is DM$_V$=14.6. The full theoretical luminosity functions extend several magnitudes fainter than the data shown in the plot.}
\label{lf_m92}
\end{figure}
Rather than adopt one or more standard filters and initial mass functions (IMF) and provide tabulated luminosity functions the distribution of isochrones comes with a program to generate luminosity functions from any of the isochrone files.  Luminosity functions may be constructed from any bandpass with a choice of bin size and IMF exponent. The IMF is assumed to be a simple power law where dN/dM $\propto$ M$^x$.  Figure \ref{lf_m92} demonstrates the luminosity functions when compared to the observed luminosity function of M92 \citep{paust}.  The isochrones on which the luminosity functions were based have $\feh$=-2.3 and $\afe$=0.4 at ages 12, 13, and 14 Gyr.  The assumed IMF is the Salpeter IMF (x=-2.35), a good fit in this case.  The distance modulus is DM$_V$=14.6 as used in Figure \ref{m92} for the Synthetic isochrone.  The theoretical luminosity functions extend approximately seven magnitudes fainter than the data in Figure \ref{m92}.

\subsection{Synthetic HB Models}

%FIGURE 6
\begin{figure}
\epsscale{0.8}
\plotone{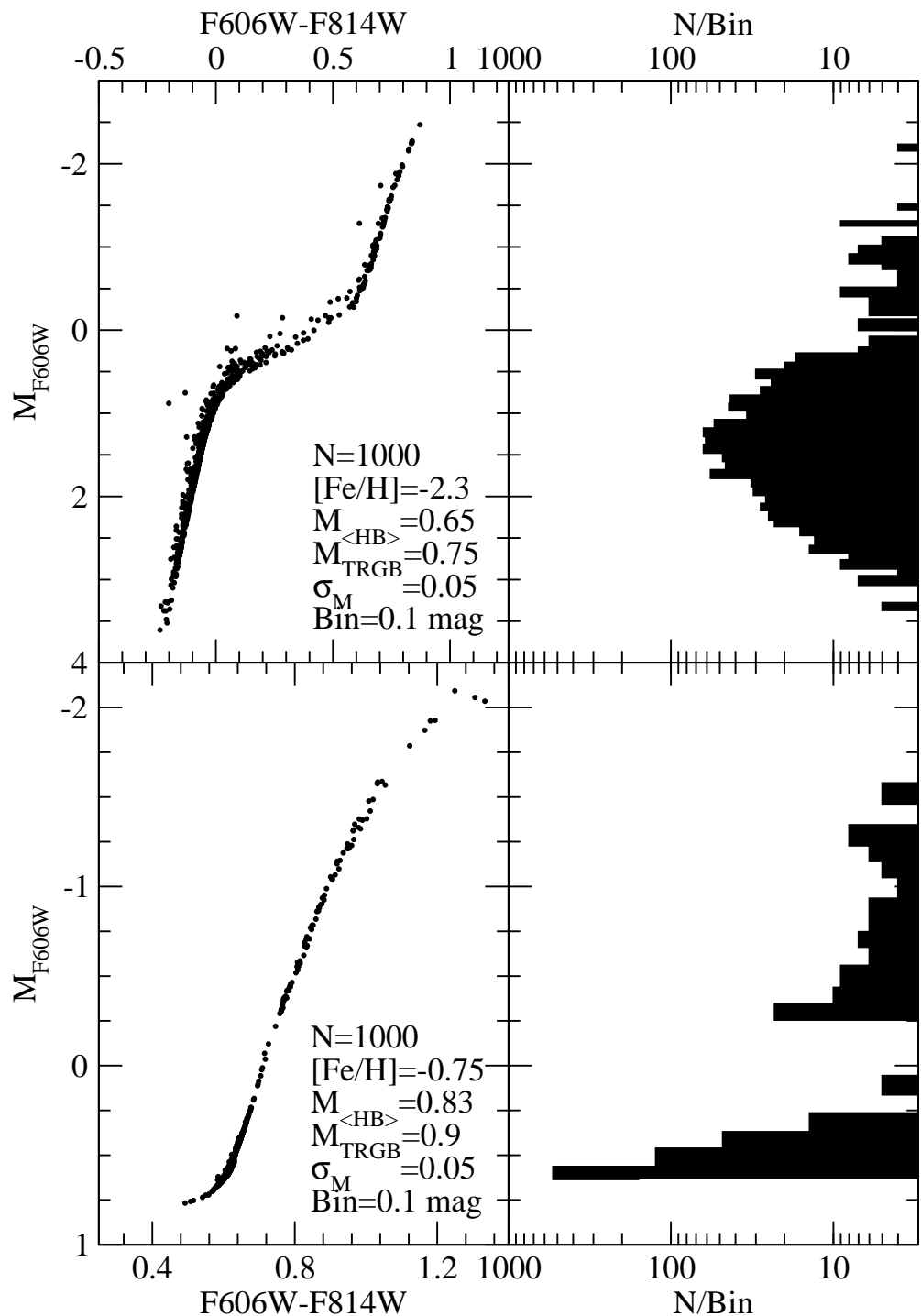}
\caption{SHB models with N=1000 stars for [Fe/H]=-2.3 (top) and -0.75 (bottom). The top panel represents a blue HB in a metal poor cluster while the bottom panel represents a red HB in a metal rich cluster.  The left hand side shows the SHB models in the color-magnitude diagram while the right hand side shows the corresponding luminosity function.}
\label{shb}
\end{figure}

The formalism for constructing synthetic horizontal branch (SHB) models is based on that of \citet{ldz}.  Accompanying the He-burning evolutionary tracks is a program for creating SHB models.  The SHB models are constructed from the tracks based on a set of parameters supplied by the user.  These parameters include the composition ($\feh$,$\afe$,Y), average HB mass ($<M_{HB}>$), the mass of the red giant branch tip (M$_{RGBT}$, effectively the upper bound to the mass distribution), the mass dispersion on the HB ($\sigma_M$), and the total number of stars (N).  Based on these parameters, the program constructs a random Gaussian distribution of stellar masses, chooses a random age, and interpolates within the tracks for mass, age, and composition.  Interpolation is linear within the tracks (age) and quadratic between the tracks  (mass and composition).  Because the actual mass distribution is finite and truncated on both ends by the finite mass range the code calculates the actual values of $<$M$_{HB}>$ and $\sigma_{HB}$ and reports these at the end.

In addition to the SHB generating program, another program is supplied to generate luminosity functions and fiducial lines from each individual SHB model.  The program works on the raw output from the SHB code by binning the SHB model data, determining the mean and standard deviation of magnitude and color in a given bin, and writing results from all bins to a file.  Use of this program in conjunction with the isochrone-based luminosity function code allows the user to create full luminosity functions that include all phases of the evolution covered by the models presented in this paper.

Figure \ref{shb} demonstrates the effect of metallicity and the mass variables mentioned above in the color-magnitude diagram for two SHB models on the right. The left side of Figure \ref{shb} shows luminosity functions constructed from the corresponding SHB model. The upper panels show the CMD and number distribution of a blue, metal poor SHB model with $\feh$=-2.3.  The lower panels show the same information for a red, metal rich SHB model with $\feh$=-0.75. These models, shown alone here, are the same used in Figures \ref{m92} and \ref{47tuc} below.

\section{Comparisons}
\subsection{Comparisons with other isochrones}
Several groups currently produce stellar evolution tracks and isochrones similar to those presented in this paper. Here the new DSEP isochrones are compared to isochrones from the other groups in both the theoretical H-R diagram and in the V vs. V--I color-magnitude diagram. Comparisons in the H-R diagram allow for differences in the luminosity and effective temperature scales to be assessed while differences in the color-magnitude diagram display additional differences due to the color-$\Teff$ transformations adopted by each group. The DSEP isochrones shown in the following figures employ the Synthetic color transformation. Figure \ref{clr1} shows the differences between Synthetic and Empirical colors.

Comparisons were performed between DSEP and BaSTI, Padova, Victoria-Regina, and Y$^2$ at 10 Gyr with Z=0.0001 assuming a scaled-solar heavy element mixture. Scaled-solar was chosen because the definition and amount of $\alpha$-enhancement is not consistent amongst the groups. Neither the primordial He abundance nor the adopted $\Delta$Y/$\Delta$Z is consistent, at Z=0.0001: BaSTI has Y=0.245, DSEP has Y=0.02452, Padova has Y=0.23, Victoria-Regina has Y=0.2352, and Y$^2$ has Y=0.2302. Aside from differences in composition, the groups do not all use the same assumptions concerning the surface boundary condition, solar-calibrated mixing length, equation of state, gravitational settling, mass loss, and other physics: each of these discrepancies will result in differences amongst the isochrones at constant age and Z.

\begin{figure}
%\epsscale{0.7}
\plotone{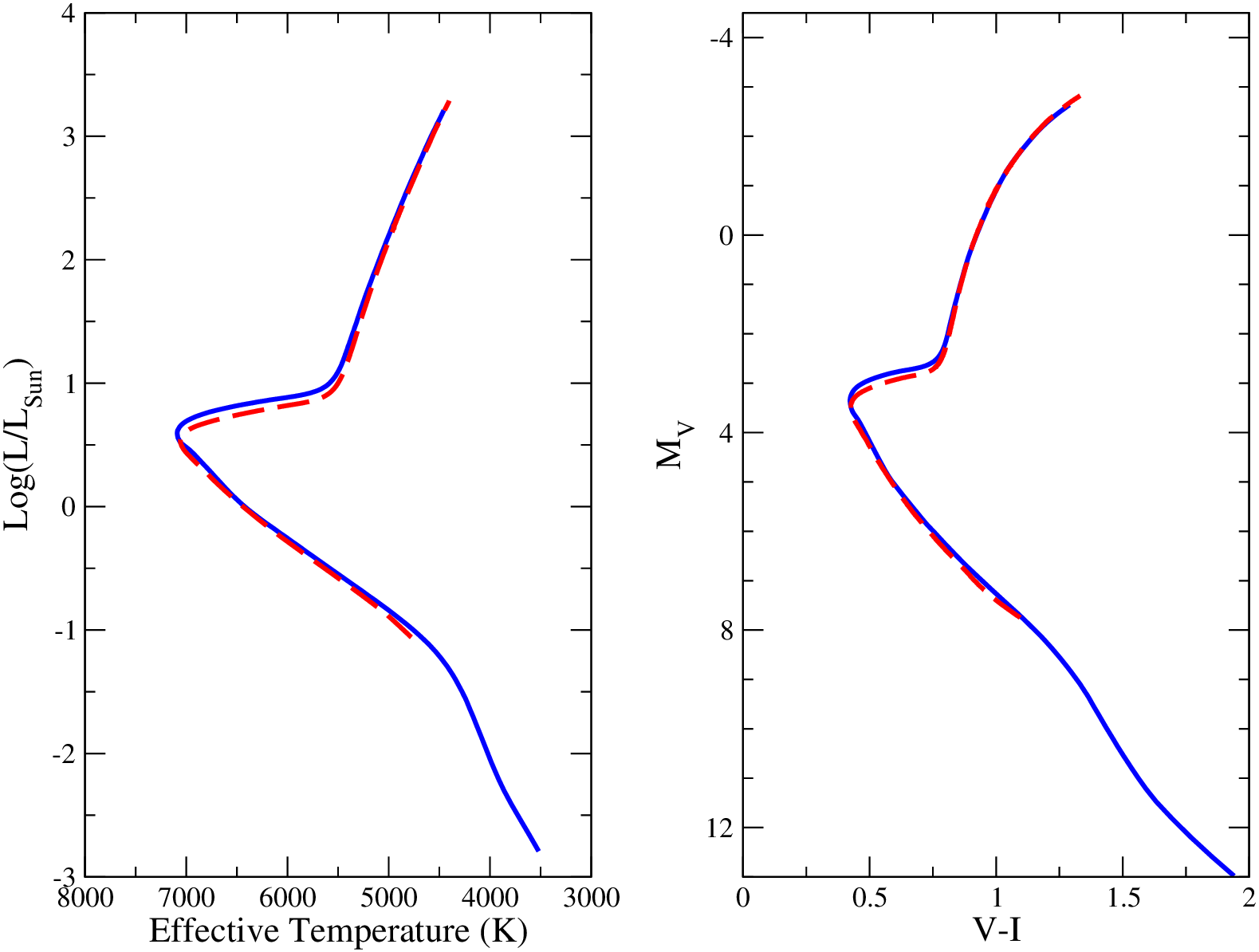}
\caption{BaSTI (dashed line) and DSEP (solid line, with Synthetic color transformation) isochrones at 10 Gyr with Z=0.0001 (scaled-solar mixture).}
\label{iso_basti}
\end{figure}

\begin{figure}
%\epsscale{0.7}
\plotone{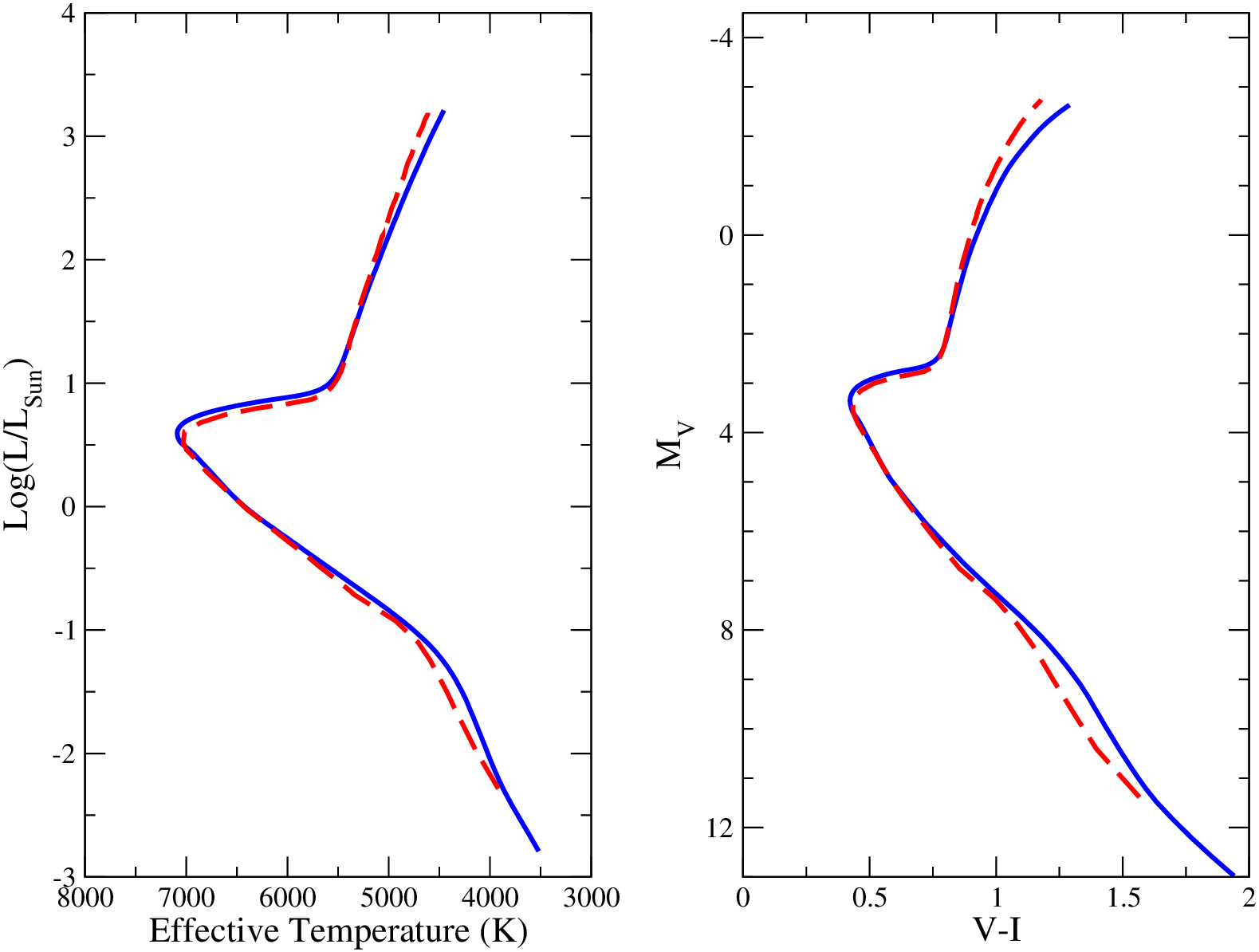}
\caption{Same as Figure \ref{iso_basti} but comparing Padova and DSEP isochrones.}
\label{iso_padova}
\end{figure}

\begin{figure}
%\epsscale{0.7}
\plotone{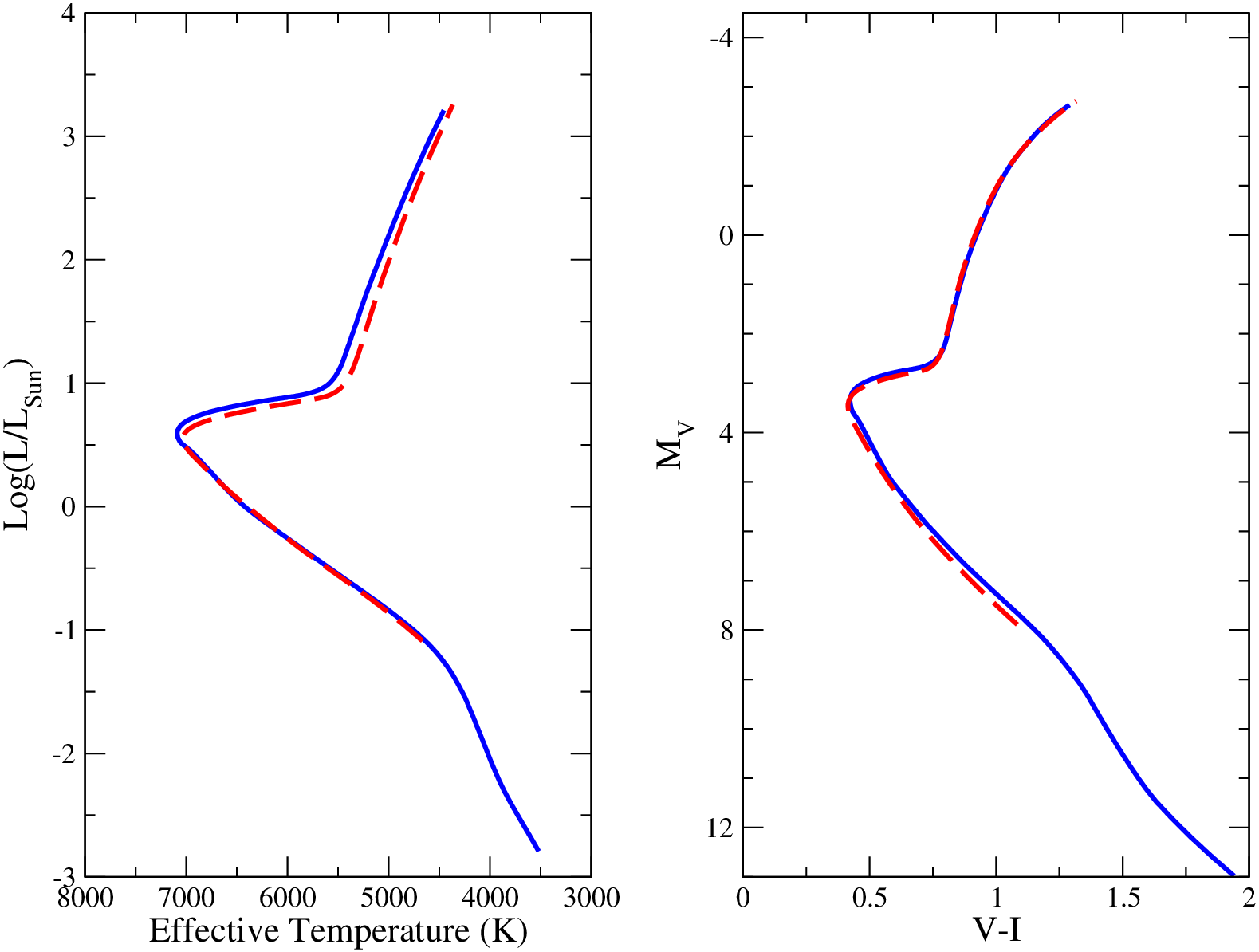}
\caption{Same as Figure \ref{iso_basti} but comparing Victoria-Regina and DSEP isochrones.}
\label{iso_victoria}
\end{figure}

\begin{figure}
%\epsscale{0.7}
\plotone{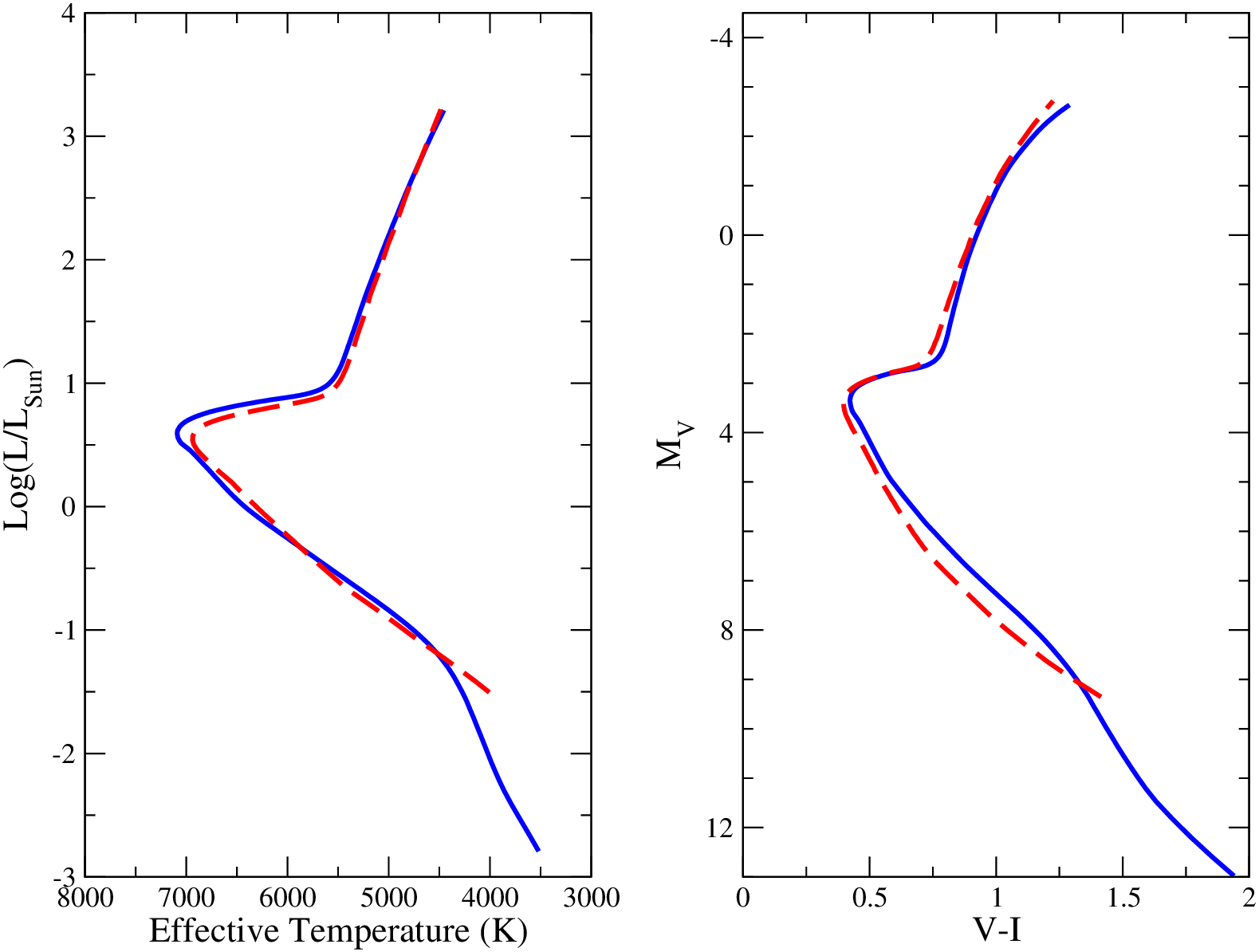}
\caption{Same as Figure \ref{iso_basti} but comparing Y$^2$ and DSEP isochrones.}
\label{iso_yale}
\end{figure}

Figure \ref{iso_basti} compares DSEP to BaSTI, Figure \ref{iso_padova} to Padova, Figure \ref{iso_victoria} to Victoria-Regina, and Figure \ref{iso_yale} to Y$^2$. Both BaSTI and Padova include core He-burning evolution in their isochrones but these portions have been removed from the figures for clarity.

In the H-R diagram, the DSEP isochrone is slightly hotter and more luminous than the others around the main sequence turn off and sub-giant branch but this does not carry over to the color-magnitude diagram in all cases due to the differences in adopted color-$\Teff$ transformations. There is good agreement on the red giant branch tip location in the H-R diagram except with Padova which is hotter than all the others.

\clearpage

%FIGURE 7

%FIGURE 8

%FIGURE 9

%FIGURE 10

\subsection{Comparisons with Data}

%FIGURE 11
\begin{figure}
%\epsscale{1.0}
\plotone{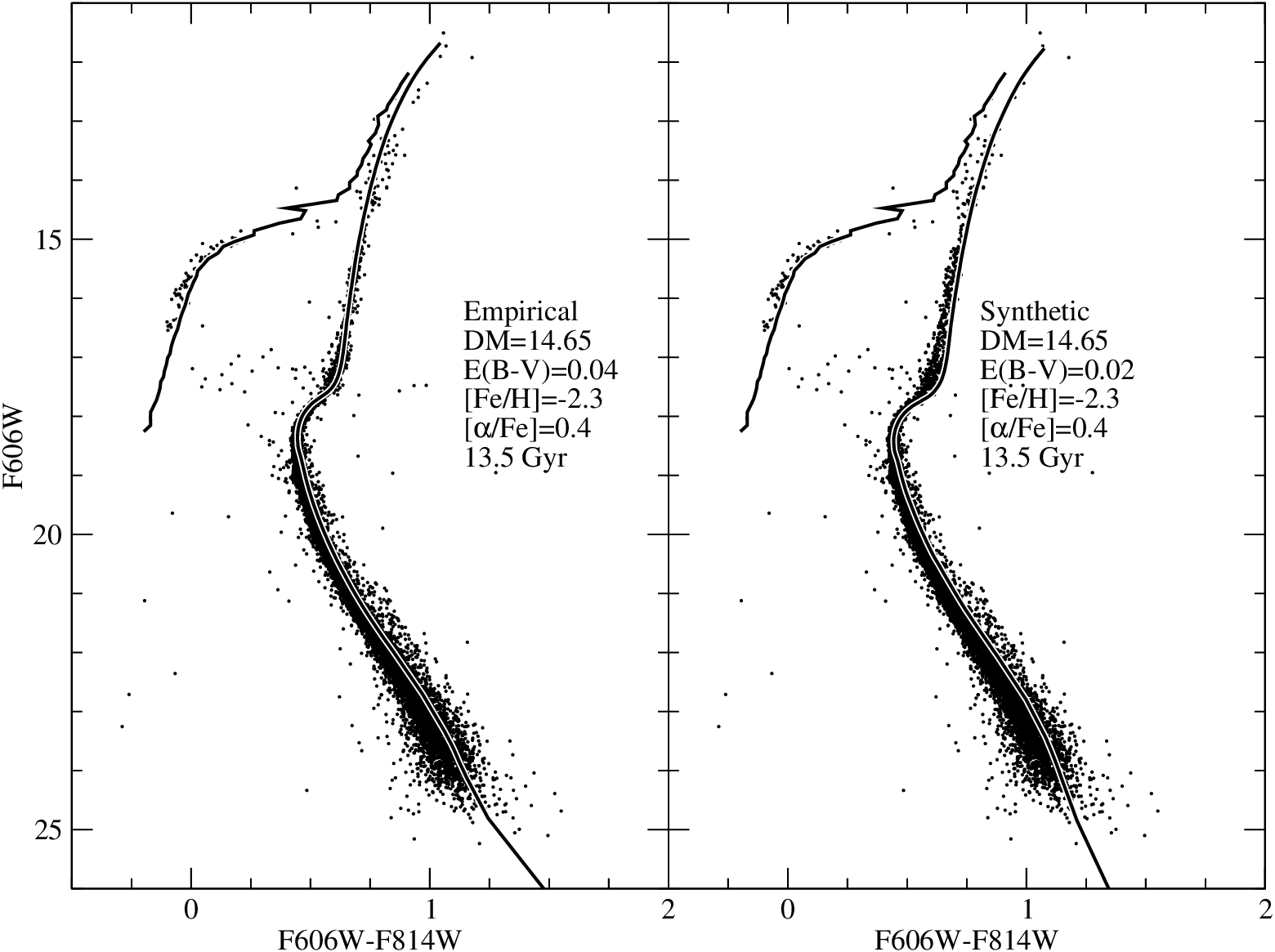}
\caption{ACS data from M92 compared to isochrones with both the Empirical (left panel) and Synthetic (right panel) color transformations. Details are listed on each panel. Data from \citet{sar}. The fiducial line from the metal poor synthetic horizontal branch model of $\S$4.4, Figure \ref{shb}, is plotted alongside both isochrones.}
\label{m92}
\end{figure}

%FIGURE 12
\begin{figure}
%\epsscale{1.0}
\plotone{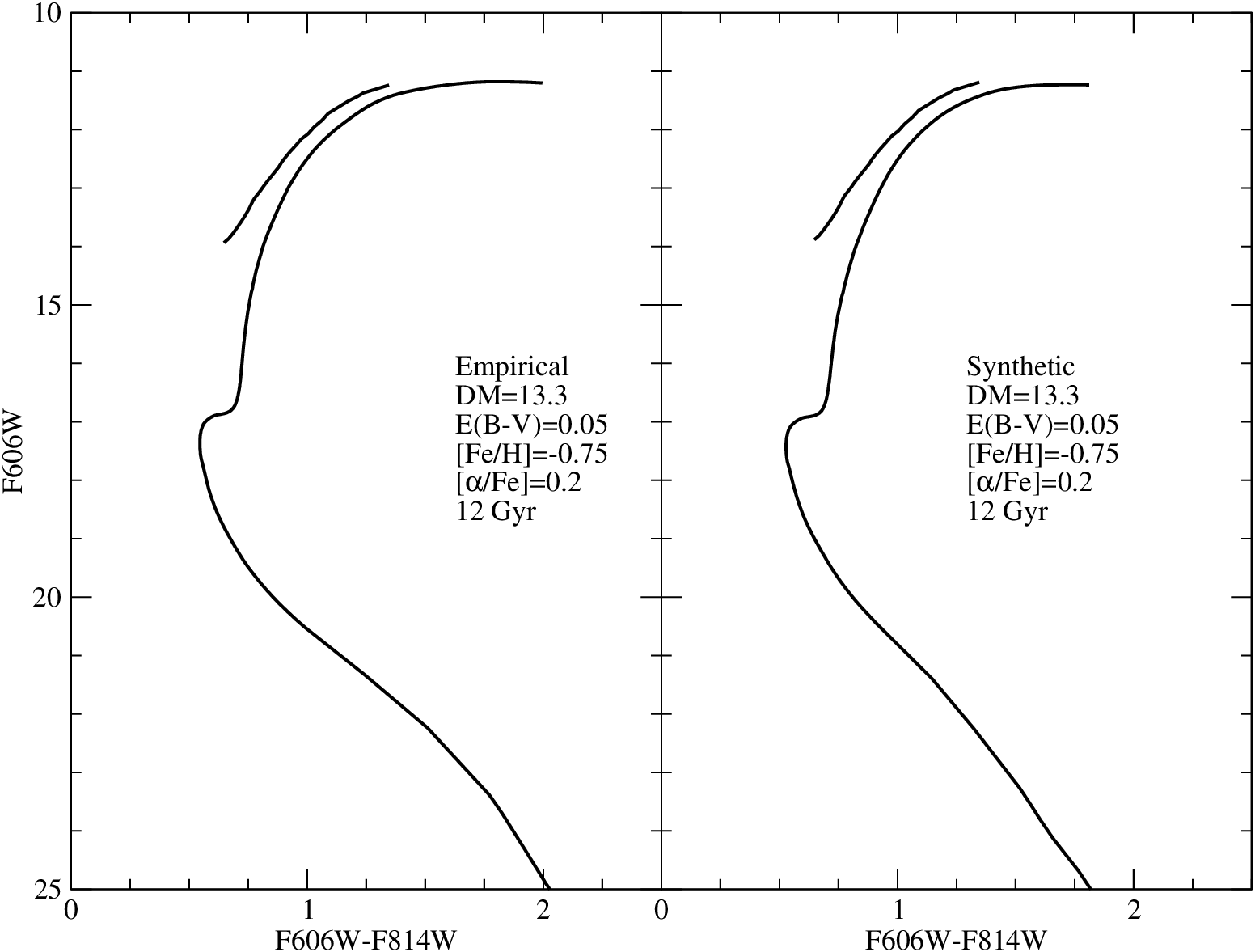}
 \caption{ACS data from 47 Tuc compared to isochrones with both the Empirical (left panel) and Synthetic (right panel) color transformations. Details are listed on each panel. Data from \citet{sar}. The fiducial line from the metal rich synthetic horizontal branch model of $\S$4.4, Figure \ref{shb}, is plotted alongside both isochrones.}
\label{47tuc}
\end{figure}

ACS photometry of two classic globular clusters, M92 and 47 Tuc, from \citet{sar} and Anderson et al. (2007, in preparation) that span a wide range of $\feh$ have been chosen for comparison.  Many of the brightest giants in Figures \ref{m92} and \ref{47tuc} include at least one saturated pixel. For these stars, the point spread function photometry used for unsaturated stars has been replaced by aperture photometry with excellent results. Full details of the photometry will be given by Anderson et al. (2007, in preparation).

Isochrones for reasonable assumptions of the age and metallicity of M 92 and 47 Tuc are shown in Figures \ref{m92} and \ref{47tuc}, respectively.  The purpose of these plots is not to constrain any of these parameters for either cluster but merely to demonstrate the performance of the isochrones when compared with high quality photometry. To show the complete evolutionary picture, ridge lines from the SHB models shown in Figure \ref{shb} are plotted alongside the photometry: the metal poor SHB model ($\feh$=-2.3) with M 92 and the metal rich SHB model ($\feh$=-0.75) with 47 Tuc assuming the adopted reddening and distance modulus in each figure.

Figure \ref{m92} shows little difference between the color transformations and both provide reasonable fits over the extent of the data. The largest difference occurs on the subgiant branch: in this region the Empirical color transformation is clearly a better fit to the data.  The behavior of the Synthetic transformation is common to most color transformations derived synthetic spectra.

Figure \ref{47tuc} shows more clearly the difference between the two color transformations.  The Empirical transformation provides a good fit from the red giant branch to approximately 3 mag below the main sequence turn off. Below that point, the Empirical isochrone becomes increasingly redder than the data. In contrast, the Synthetic isochrone becomes bluer than much of the data a few magnitudes below the turn off. The difference between the Synthetic isochrone and the data is smaller than in the case of the Empirical isochrone, and in the opposite sense, but still visible.  For this reason, the Synthetic color transformation is preferable for examining the lower main sequence, particularly at higher metallicity. Disagreement between the lower main sequence data and the isochrones is likely due to errors in both the effective temperature scale of the stellar evolution models and in the color transformations but it is not possible to quantify these errors.

\section{Conclusions}
The stellar evolution code DSEP has been modified to significantly improve the accuracy of isochrones on the lower main sequence and to self-consistently treat the outer layers of stars for a wide range of compositions.  The synthetic and empirical color transformations employed have strengths and weaknesses and highlight the amount of uncertainty in current color-$\Teff$ transformations in the color-magnitude diagram.

A new set of stellar evolution models have been computed with DSEP for a range of initial He abundance, $\feh$, and $\afe$ encompassing ages and abundances observed in Galactic globular clusters.  The models are presented as isochrones (2-15 Gyr) and stellar evolution tracks (0.1-1.8 $\Ms$) with accompanying programs to create luminosity functions and SHB models suitable for analysis of old stellar populations.  Details of the stellar evolution calculations and all steps leading up to the final products have been discussed.  These models should prove useful in analyzing the ACS Galactic Globular Cluster Survey data and in many other applications.  The models extend to 0.1 $\Ms$ on the main sequence, making them useful for deep photometric studies such as the ACS Galactic Globular Cluster Survey and at the same time provide nearly full evolutionary coverage of the stars found in globular clusters and old open clusters.

\acknowledgments

AD wishes to thank: Alan Irwin for his work on FreeEOS and for making it freely available; Alan Boothroyd for sharing his subroutines to manage the OPAL Type 2 opacity tables; and Allen Sweigart and Achim Weiss for helpful advice regarding HB and AGB evolution.

The work of DJ and EB was supported in part by by NASA grants NAG5-3505 and NAG5-12127, and NSF grant AST-0307323. This research used resources of the National Energy Research Scientific Computing Center (NERSC), which is supported by the Office of Science of the U.S. Department of Energy under Contract No. DE-AC03-76SF00098; and the H\"ochstleistungs Rechenzentrum Nord (HLRN).  We thank both these institutions for a generous allocation of computer time.

JF acknowledges support from NSF grant AST-0239590 and Grant No. EIA-0216178, Grant No. EPS-0236913 with matching support from the State of Kansas and the Wichita State University High Performance Computing Center.

Support for this work (GO-10775) was provided by NASA through a grant from the Space Telescope Science Institute which is operated by the Association of Universities for Research in Astronomy, Incorporated, under NASA contract NAS5-26555.

\begin{appendix}
\section{Composition, Color Transformations, and File Names}
Composition is specified in the file names in terms of $\feh$ as \texttt{feh} and $\afe$ as \texttt{afe}.  The values of $\feh$ and $\afe$ are listed using the letters \texttt{m} for minus ($<$0) and \texttt{p} for plus ($>$0) followed by two digits for $\feh$ and one digit for $\afe$.  For example, $\feh$=-2.5, $\afe$=+0.4 would be represented by the string \texttt{fehm25afep4}. Zero is set to positive and thus $\feh$=0, $\afe$=0 is listed as \texttt{fehp00afep0}.  Tracks and isochrones with enhanced He, either Y=0.33 or Y=0.4, have an additional string following \texttt{fehXXXafeXX} in their filenames: \texttt{y33} or \texttt{y40}, respectively. 

Color transformations are listed as a suffix to isochrone and evolutionary track file names.  The synthetic color transformation is indicated by the suffix \texttt{.phx} and the semi-empirical color transformation by \texttt{.cmd}. As a final example, the synthetic color isochone with $\feh$=-1.5, $\afe$=+0.4, and Y=0.33 is called \texttt{fehm15afep4y33.phx}.

\section{Stellar Evolution Track Files}
The stellar evolution tracks begin on the pre-main sequence and terminate at the He flash or begin on the ZAHB and terminate at the onset of thermal pulsations on the AGB (TP-AGB).  Tracks are provided for masses ranging from 0.1 to 1.8 $\Ms$ in increments of 0.05 $\Ms$. For the mass range that experiences the He flash, separate He-burning tracks are provided. The content of these tracks are identical to the H-burning tracks as described below.

The track files come in two formats: one containing theoretical quantities and one containing observable quantities.  The theoretical track files contain information about the initial composition and physics used in a header followed by data lines.  Each data line lists age in years, log $\Teff$, log g, log L/$\Ls$, log R/$\Rs$, Y$_{core}$, Z$_{core}$, (Z/X)$_{surf}$, L$_H$/$\Ls$, L$_{He}$/$\Ls$, M$_{He-core}$, and M$_{CO-core}$. The quantity L$_H$ is the total H-fusion luminosity combining all branches of the p-p chain and CNO cycle luminosities while L$_{He}$ combines luminosities from the triple-$\alpha$ and $^{12}$C($\alpha$,$\gamma$)$^{16}$O reactions.

The naming convention for the stellar evolution tracks is to use the prefix \texttt{m} (for mass) followed by three integers representing the stellar mass (1 $\Ms$ $\rightarrow$ \texttt{m100}) followed by composition information and either the suffix \texttt{.trk} for theoretical quantities or the appropriate suffix for a color transformation specified above. The He-burning tracks use the same naming convention except the \texttt{m} has been replaced with \texttt{hb}.

The header and first eight data lines of the 1 $\Ms$, $\feh$=0, $\afe$=0, Y=0.27 stellar evolution track are given below for both the theoretical file \texttt{m100fehp00afep0.trk} in Table \ref{trk1} and the observable file with the PHOENIX color transformation \texttt{m100fehp00afep0.phx} in Table \ref{trk2}. The columns labeled F606W and F814W refer to the ACS bands and the columns labeled F6062 and F8142 refer to the WFPC2 bands.

\begin{deluxetable}{cccccccccccc}
\rotate
\tabletypesize{\footnotesize}
\tablecolumns{12}
\tablewidth{0pc}
\tablecaption{Example of a Stellar Evolution Track with Theoretical Quantities\label{trk1}}
\tablehead{\multicolumn{12}{l}{M=1.00 X=7.0710E-01 Z=1.8850E-02 COMP=GS98 [a/Fe]= 0.00 MIXL=1.9380 CCO=0.00}\\
\colhead{Age (yrs)}&\colhead{Log T}&\colhead{Log g}&\colhead{Log L}&\colhead{Log R}&\colhead{Y\_core}&\colhead{Z\_core}&\colhead{(Z/X)\_surf}&\colhead{L\_H}&\colhead{L\_He}&\colhead{M\_He\_core}&\colhead{M\_CO\_core}}

\startdata
5.27000000E+02 & 3.62873 & 2.41053 & 1.49574 & 1.01367 &2.7402E-01 &1.8850E-02 &2.6658E-02 & 0.0000E+00 & 0.0000E+00& 0.0000E+00 &0.0000E+00\\
5.94200000E+02 & 3.62878 & 2.41191 & 1.49456 & 1.01298 &2.7402E-01 &1.8850E-02 &2.6658E-02 & 0.0000E+00 & 0.0000E+00& 0.0000E+00 &0.0000E+00\\
6.64760000E+02 & 3.62884 & 2.41335 & 1.49335 & 1.01225 &2.7402E-01 &1.8850E-02 &2.6658E-02 & 0.0000E+00 & 0.0000E+00& 0.0000E+00 &0.0000E+00\\
7.38848000E+02 & 3.62890 & 2.41487 & 1.49209 & 1.01150 &2.7402E-01 &1.8850E-02 &2.6658E-02 & 0.0000E+00 & 0.0000E+00& 0.0000E+00 &0.0000E+00\\
8.16640400E+02 & 3.62897 & 2.41645 & 1.49077 & 1.01071 &2.7402E-01 &1.8850E-02 &2.6658E-02 & 0.0000E+00 & 0.0000E+00& 0.0000E+00 &0.0000E+00\\
8.98322420E+02 & 3.62904 & 2.41810 & 1.48940 & 1.00988 &2.7402E-01 &1.8850E-02 &2.6658E-02 & 0.0000E+00 & 0.0000E+00& 0.0000E+00 &0.0000E+00\\
9.84088541E+02 & 3.62911 & 2.41983 & 1.48796 & 1.00902 &2.7402E-01 &1.8850E-02 &2.6658E-02 & 0.0000E+00 & 0.0000E+00& 0.0000E+00 &0.0000E+00\\
1.07414297E+03 & 3.62919 & 2.42163 & 1.48646 & 1.00811 &2.7402E-01 &1.8850E-02 &2.6658E-02 & 0.0000E+00 & 0.0000E+00& 0.0000E+00 &0.0000E+00
\enddata
\end{deluxetable}

\begin{deluxetable}{ccccccccccc}
\rotate
\tabletypesize{\footnotesize}
\tablecolumns{11}
\tablewidth{0pc}
\tablecaption{Example of a Stellar Evolution Track with Observable Quantities\label{trk2}}
\tablehead{\multicolumn{11}{l}{M/Mo=1.00 [Fe/H]= 0.07 [a/Fe]= 0.00 Y=0.274}\\
\colhead{Age (yrs)}&\colhead{Log T}&\colhead{Log g}&\colhead{Log L}&\colhead{B}&\colhead{V}&\colhead{I}&\colhead{F606W}&\colhead{F814W}&\colhead{F6062}&\colhead{F8142}}
\startdata
5.27000000E+02 &3.62873 & 2.41053 & 1.49574 & 3.1495 & 1.7757 & 0.3959 & 1.4607 & 0.4112 & 1.3968 & 0.4286\\
5.94200000E+02 &3.62878 & 2.41191 & 1.49456 & 3.1517 & 1.7781 & 0.3988 & 1.4632 & 0.4140 & 1.3994 & 0.4315\\
6.64760000E+02 &3.62884 & 2.41335 & 1.49335 & 3.1539 & 1.7806 & 0.4017 & 1.4657 & 0.4169 & 1.4019 & 0.4344\\
7.38848000E+02 &3.62890 & 2.41487 & 1.49209 & 3.1562 & 1.7832 & 0.4047 & 1.4684 & 0.4199 & 1.4046 & 0.4374\\
8.16640400E+02 &3.62897 & 2.41645 & 1.49077 & 3.1585 & 1.7859 & 0.4079 & 1.4712 & 0.4231 & 1.4074 & 0.4405\\
8.98322420E+02 &3.62904 & 2.41810 & 1.48940 & 3.1609 & 1.7886 & 0.4112 & 1.4740 & 0.4263 & 1.4102 & 0.4438\\
9.84088541E+02 &3.62911 & 2.41983 & 1.48796 & 3.1635 & 1.7916 & 0.4146 & 1.4771 & 0.4298 & 1.4133 & 0.4472\\
1.07414297E+03 &3.62919 & 2.42163 & 1.48646 & 3.1662 & 1.7946 & 0.4182 & 1.4802 & 0.4334 & 1.4164 & 0.4508
\enddata
\end{deluxetable}
 
\section{Isochrones and Luminosity Functions}
Isochrone files begin with a header that lists composition and color transformation details as well as the number of ages in the file.  The isochrone files distributed with this paper contain 27 ages from 2 to 15 Gyr in increments of 0.5 Gyr. The individual isochrones have their own headers consisting of two lines.  The first line gives the age of the isochrone and the number of data points (EEPs) it contains.  The second line lists the quantities provided on each line: EEP number, stellar mass, log $\Teff$, log g, log L/$\Ls$, and the absolute magnitudes in all seven filters.  The header and first six data lines of the $\feh$=0, $\afe$=0 isochrone file \texttt{fehp00afep0.cmd} are printed in Table \ref{isotab}.  The columns labeled F606W and F814W refer to the ACS bands and the columns labeled F6062 and F8142 refer to the WFPC2 bands.

\begin{deluxetable}{cccccccccccc}
\rotate
\tabletypesize{\footnotesize}
\tablecolumns{12}
\tablewidth{0pc}
\tablecaption{Example of an Isochrone File\label{isotab}}
\tablehead{
\multicolumn{12}{l}{NUMBER OF AGES=29}\\
\cline{1-6}\\
\colhead{MIX-LEN}&\colhead{Y}&\colhead{Z}&\colhead{Zeff}&\colhead{[Fe/H]}&\colhead{[a/Fe]}&\multicolumn{6}{c}{}\\
\colhead{1.9380}&\colhead{0.2741}&\colhead{1.8850E-02}&\colhead{1.8850E-02}&\colhead{0.07}&\colhead{0.00}&\multicolumn{6}{c}{}\\
\cline{1-6}\\
\multicolumn{12}{l}{\*\*PHOTOMETRIC SYSTEM\*\*: V\&C(2003)\tablenotemark{a}+Sirianni et al.(2005)}\\
\cline{1-6}\\
\multicolumn{12}{l}{AGE= 2.000 EEPS=238}\\
\colhead{EEP}&\colhead{M/Mo}&\colhead{LogTeff}&\colhead{LogG}&\colhead{LogL/Lo}&\colhead{B}&\colhead{V}&\colhead{I}&\colhead{F606W}&\colhead{F814W}&\colhead{F6062}&\colhead{F8142}}
\startdata
   4 & 0.099162 & 3.5048 & 5.3118 &-2.9048 &16.5676 &14.8000 &11.7423 &14.1827 &11.6499 &14.0308 &11.6953\\
   5 & 0.115037 & 3.5048 & 5.2473 &-2.7758 &16.2406 &14.4760 &11.4311 &13.8600 &11.3397 &13.7098 &11.3849\\
   6 & 0.134056 & 3.5048 & 5.1810 &-2.6432 &15.9058 &14.1439 &11.1106 &13.5292 &11.0200 &13.3805 &11.0652\\
   7 & 0.169755 & 3.5076 & 5.1016 &-2.4501 &15.3348 &13.5914 &10.6076 &12.9822 &10.5209 &12.8399 &10.5654\\
   8 & 0.223896 & 3.5162 & 5.0395 &-2.2333 &14.5387 &12.8438 & 9.9844 &12.2493 & 9.9068 &12.1222 & 9.9497\\
   9 & 0.276573 & 3.5236 & 4.9959 &-2.0684 &13.9167 &12.2621 & 9.5074 &11.6808 & 9.4371 &11.5657 & 9.4787
\enddata
\tablenotetext{a}{V\&C2003 refers to \citet{vdb}.}
\end{deluxetable}

Output from the luminosity function program (\texttt{isolf.f}) contains the same header as the isochrone file it was created from.  Likewise, each luminosity function has a two line header that lists the age and number of data points followed by a description of the information printed in each line.  The data lines list a bin number,  the average absolute magnitude of the bin, the average color of the bin, the cumulative (logged) number of stars, and the (logged) number of stars in that bin. The cumulative star count begins with the brightest bin. The header and first six data lines of a luminosity function generated from the isochrone file (Table \ref{isotab}) are printed in Table \ref{lftab}.  The luminosity function is created using the F606W (ACS) filter with color F606W-F814W, bin size of 0.1 mag, and Salpeter IMF.

\begin{deluxetable}{cccccc}
\rotate
\tabletypesize{\footnotesize}
\tablecolumns{6}
\tablewidth{0pc}
\tablecaption{Example of a Luminosity Function File\label{lftab}}
\tablehead{
\multicolumn{6}{l}{NUMBER OF AGES=29}\\
\cline{1-6}\\
\colhead{MIX-LEN}&\colhead{Y}&\colhead{Z}&\colhead{Zeff}&\colhead{[Fe/H]}&\colhead{[a/Fe]}\\
\colhead{1.9380}&\colhead{0.2741}&\colhead{1.8850E-02}&\colhead{1.8850E-02}&\colhead{0.07}&\colhead{0.00}\\
\cline{1-6}\\
\multicolumn{6}{l}{\*\*PHOTOMETRIC SYSTEM\*\*: V\&C(2003)+Sirianni et al.(2005)}\\
\cline{1-6}\\
\multicolumn{6}{l}{AGE= 2.000 BINS=158 IMF x= -2.35}\\
\colhead{N}&\colhead{MF606W}&\colhead{F606W-F814W}&\colhead{Log10(N)}&\colhead{Log10(dN)}&\colhead{}}
\startdata
  1 & -1.517  &   1.7545    &     0.1086    &     0.1086& \phn \\ 
  2 & -1.417  &   1.7064    &     0.3138    &    -0.1105& \phn \\ 
  3 & -1.317  &   1.4011    &     0.4090    &    -0.2966& \phn \\ 
  4 & -1.217  &   1.3380    &     0.4921    &    -0.2671& \phn \\ 
  5 & -1.117  &   1.2829    &     0.5655    &    -0.2426& \phn \\ 
  6 & -1.017  &   1.2381    &     0.6276    &    -0.2478& \phn    
\enddata
\end{deluxetable}

\section{Synthetic Horizontal Branch Models and Luminosity Functions}
Output from the SHB code (\texttt{shb.f}) consists of a file containing one line for each star plus a short header. The header lists number of stars, $\feh$, $\afe$, and Y$_{init}$.  The data lines include stellar mass, age, log $\Teff$, log g, log L/$\Ls$, and absolute magnitudes in all seven filters.  Table \ref{shbtab} shows the header and first eight data lines from an SHB model with 1000 stars at $\feh$=0, $\afe$=0, and Y=0.274.

\begin{deluxetable}{cccccccccccc}
\rotate
\tabletypesize{\footnotesize}
\tablecolumns{12}
\tablewidth{0pc}
\tablecaption{Example of a Synthetic Horizontal Branch File\label{shbtab}}
\tablehead{
\multicolumn{12}{l}{N=    1000  [Fe/H]= 0.00  [a/Fe]=0.00  Y\_i=0.274}\\
\colhead{M/Mo}&\colhead{Age (yrs)}&\colhead{Log Teff}&\colhead{Log g}&\colhead{Log L/Lo}&\colhead{B}&\colhead{V}&\colhead{I}&\colhead{F606W}&\colhead{F814W}&\colhead{F6062}&\colhead{F8142} 
}
\startdata
  0.7771 & 1.0639E+08 & 3.6677 & 2.2732 & 1.6791 & 2.1224 & 0.9687 &-0.1202 & 0.7106 &-0.1177 & 0.7032 &-0.1000\\
  0.7920 & 9.7970E+07 & 3.6727 & 2.3536 & 1.6274 & 2.1900 & 1.0672 & 0.0076 & 0.8166 & 0.0106 & 0.8100 & 0.0279\\
  0.8825 & 6.9989E+07 & 3.6760 & 2.4404 & 1.6004 & 2.2174 & 1.1160 & 0.0741 & 0.8700 & 0.0774 & 0.8639 & 0.0944\\
  0.7812 & 5.2351E+07 & 3.6792 & 2.4334 & 1.5671 & 2.2678 & 1.1813 & 0.1560 & 0.9397 & 0.1596 & 0.9339 & 0.1763\\
  0.8525 & 1.0007E+08 & 3.6694 & 2.3350 & 1.6638 & 2.1377 & 0.9968 &-0.0823 & 0.7412 &-0.0796 & 0.7341 &-0.0620\\
  0.8959 & 1.1958E+08 & 3.6137 & 1.5453 & 2.2529 & 1.4486 &-0.0369 &-1.5517 &-0.3977 &-1.5599 &-0.4197 &-1.5359\\
  0.7885 & 9.2760E+07 & 3.6747 & 2.3783 & 1.6087 & 2.2134 & 1.1021 & 0.0536 & 0.8544 & 0.0568 & 0.8480 & 0.0739\\
  0.8229 & 1.0110E+08 & 3.6695 & 2.3239 & 1.6602 & 2.1459 & 1.0050 &-0.0734 & 0.7495 &-0.0707 & 0.7424 &-0.0532
\enddata
\end{deluxetable}

Output from \texttt{shb.f} may be processed by \texttt{shblf.f} to create a fiducial line and luminosity function from the SHB model. The \texttt{shblf} program creates luminosity functions and fiducials in any of the seven bands and with either a set number of bins or set bin size.  It scans through and SHB model file and calculates the number of stars, the average and standard deviation of the absolute magnitude, and the average and standard deviation of the color in each bin and finally prints these data along with a short header. Table \ref{shblftab} show the header and first eight data lines of an SHB luminosity function derived from the SHB model described above (Table \ref{shbtab}).

\begin{deluxetable}{cccccc}
\rotate
\tabletypesize{\footnotesize}
\tablecolumns{6}
\tablewidth{0pc}
\tablecaption{Example of a Synthetic Horizontal Branch Luminosity Function File\label{shblftab}}
\tablehead{
\multicolumn{6}{l}{N=    1000  [Fe/H]= 0.00  [a/Fe]=0.00  Y\_i=0.274}\\
\multicolumn{6}{l}{Bin Size= 0.0368 Number of Bins=  50}\\
\colhead{N/Bin}&\colhead{Bin Center}&\colhead{$<$F606-F814$>$}&\colhead{sigma}&\colhead{$<$F606W$>$}&\colhead{sigma}
}    
\startdata
    5 &-0.82520300 & 1.62138000 & 0.11453257 &-0.82572000 & 0.01088511\\
    2 &-0.78840900 & 1.65370000 & 0.11840000 &-0.79715000 & 0.00715000\\
    4 &-0.75161500 & 1.69665000 & 0.24597608 &-0.74417500 & 0.00733941\\
    3 &-0.71482100 & 1.41780000 & 0.02011765 &-0.71806667 & 0.00549201\\
    2 &-0.67802700 & 1.36825000 & 0.03005000 &-0.68070000 & 0.01010000\\
    2 &-0.64123300 & 1.66125000 & 0.34805000 &-0.64785000 & 0.01055000\\
    3 &-0.60443900 & 1.55713333 & 0.35709536 &-0.60706667 & 0.00978445
\enddata                                               
\end{deluxetable}

\section{Computer Programs and Summary}
Five computer programs, written in Fortran 77 are available along with the data files.  These programs allow for the following manipulations of the data files: 

\begin{description}
\item \texttt{iso\_interp\_feh.f}: Isochrone interpolation in $\feh$ at fixed $\afe$, Y, and age
\item \texttt{iso\_split.f}: Create individual age isochrone files from one of the large, many-aged files
\item \texttt{isolf.f}: Luminosity function construction from an isochrone file
\item \texttt{shb.f}: Synthetic horizontal branch model construction with $\feh$ interpolation
\item \texttt{shblf.f}: Luminosity function/fiducial line construction from a SHB model file
\end{description}

An accompanying \texttt{README} text file contains the text of this appendix.

\section{How to obtain the data files and programs}
The entire distribution of stellar evolution tracks, isochrones, and accompanying computer programs may be downloaded from \url{http://stellar.dartmouth.edu/$\sim$models/} and the Multimission Archive at the Space Telescope Science Institute (MAST; \url{http://archive.stsci.edu/}) as part of the ACS Galactic Globular Cluster Survey (GO-10775).
\end{appendix}


\clearpage

\begin{thebibliography}{}
\bibitem[Adelberger et al.(2002)]{adel}Adelberger, E. C. et al. 2002, Rev. Mod. Phys., 70, 1265
\bibitem[Bahcall et al.(2005)]{bah}Bahcall, J. N., Basu, S., Pinsonneault, M., \& Serenelli, A. M. 2005, \apj, 618, 1049
\bibitem[Baraffe et al.(1997)]{bar}Baraffe, I., Chabrier, G., Allard, F., \& Hauschildt, P. H. 1997, \aap, 327, 1054
\bibitem[Bedin et al.(2005)]{bed}Bedin, L. R., Cassisi, S., Castelli, F., Piotto, G., Anderson, J., Salaris, M.,
Momany, Y., \& Pietrinferni, A. 2005, \mnras, 357, 1038
\bibitem[Berbusch \& Vandenberg(2001)]{ber}Bergbusch, P. A., \& VandenBerg, D. A. 2001, \apj, 556, 322
\bibitem[Bessell(1990)]{bes}Bessell, M. S. 1990, \pasp, 102, 1181
\bibitem[Bjork \& Chaboyer(2006)]{bjork}Bjork, S. R. \& Chaboyer, B. 2006, \apj, 641, 1102
\bibitem[Bonatto et al.(2004)]{bon}Bonatto, Ch., Bica, E., Girardi, L. 2004, \aap, 415, 571
\bibitem[Canuto(1970)]{canu}Canuto 1970, \apj, 159, 641
\bibitem[Castellani et al.(1971)]{cast}Castellani, V., Giannone, P., \& Renzini, A. 1971, \apss, 10, 340
\bibitem[Castelli \& Kurucz(2003)]{cas}Castelli, F., \& Kurucz, R. L. 2003, IAU Symp. 210, ed. N. Piskunov, W. W. Weiss, \& D. F. Gray (San Francisco: ASP), A20
\bibitem[Chaboyer et al.(2001)]{cha}Chaboyer, B., Fenton, W. H., Nelan, J. E., Patnaude, D. J., \& Simon, F. E. 2001, \apj, 562, 521
\bibitem[Chaboyer \& Kim(1995)]{cha2}Chaboyer, B. \& Kim, Y.-C. 1995, \apj, 454, 767
\bibitem[Cordier et al.(2007)]{cord}Cordier, D., Pietrinferni, A., Cassisi, S., \& Salaris, M. 2007, \aj, 133, 468
\bibitem[Demarque et al.(2004)]{dem}Demarque, P., Woo, J.-H., Kim, Y.-C., \& Yi, S. K. 2004, \apjs, 155, 667
\bibitem[Dorman \& Rood(1993)]{dorm}Dorman, B. \& Rood, R. T. 1993, \apj, 409, 387
\bibitem[Ferguson et al.(2005)]{fer}Ferguson, J. W., Alexander, D. R., Allard, F., Barman, T., Bodnarik, J. G., Hauschildt, P. H., Heffner-Wong, A., \& Tamanai, A. 2005 \apj, 623, 585
\bibitem[Girardi et al.(2002)]{gir}Girardi, L., Bertelli, G., Bressan, A., Chiosi, C., Groenewegen, M. A. T., Marigo, P., Salasnich, B., \& Weiss, A. 2002, \aap, 391, 195
\bibitem[Girardi et al.(2004)]{gir2}Girardi, L., Grebel, E. K., Odenkirchen, M., Chiosi, C. 2004, \aap, 422, 205
\bibitem[Gratton et al.(2001)]{grat}Gratton, R. G., et al. 2001, \aap, 369, 87
\bibitem[Grevesse \& Sauval(1998)]{gre}Grevesse, N. \& Sauval, A. J. 1998, \ssr, 85, 161
\bibitem[Haft et al.(1994)]{haft}Haft, M., Raffelt, G., \& Weiss, A. 1994, \apj, 425, 222
\bibitem[Hauschildt et al.(1999a)]{phxa}Hauschildt, P. H., Allard, F., \& Baron, E. 1999a, \apj, 512, 377 
\bibitem[Hauschildt et al.(1999b)]{phxb}Hauschildt, P. H., Allard, F., Ferguson, J., Baron, E., \& Alexander, D. 1999b, \apj, 525, 871
\bibitem[Hubbard \& Lampe(1969)]{hubb}Hubbard, W. B. \& Lampe, M. 1969, \apjs, 18, 297
\bibitem[Iglesias \& Rogers(1996)]{igl}Iglesias, C.A. \& Rogers, F.J. 1996, \apj, 464, 943
\bibitem[Imbriani et al.(2004)]{imbr}Imbriani, G. et al. 2004, \aap, 420, 625
\bibitem[Irwin(2004)]{irwin} Irwin, A.  2004, Technical Report (http://freeeos.sourceforge.net)
\bibitem[Korn et al.(2006)]{korn}Korn, A.J., Grundahl, F., Richard, O., Barklem, P.S., Mashonkina, L., Collet, R., Piskunov, N., \& Gustafsson, B., 2006, The Messenger, 125, 6
\bibitem[Krishna Swamy(1966)]{kri}Krishna Swamy, K. S. 1966, \apj, 145, 174
\bibitem[Kunz et al.(2002)]{kunz}Kunz, R., Fey, M., Jaeger, M., Mayer, A., \& Hammer, J. W. 2002, \apj, 567, 643
\bibitem[Lee, Demarque, \& Zinn(1990)]{ldz} Lee, Y.-W., Demarque, P., \& Zinn, R. 1990, \apj, 350, 155
\bibitem[Paust et al.(2007)]{paust}Paust, N. E. Q., Chaboyer, B., \& Sarajedini, A. 2007, \aj, in press
\bibitem[Piersanti et al.(2004)]{pier}Piersanti, L., Tornambe, A., \& Castellani, V. 2004, \mnras, 353, 243
\bibitem[Pietrinferni et al.(2004)]{piet}Pietrinferni, A., Cassisi S., Salaris M. \& Castelli F. 2004, \apj, 612, 168
\bibitem[Pietrinferni et al.(2006)]{piet2}Pietrinferni, A., Cassisi, S., Salaris, M., \& Castelli, F. 2006, \apj, 642, 797
\bibitem[Ram\'irez \& Cohen(2003)]{ram}Ram\'irez, S. V. \& Cohen, J. G. 2003, \aj, 125, 224
\bibitem[Sarajedini et al.(2007)]{sar}Sarajedini, A., Bedin, L. R., Chaboyer, B., Dotter, A., Siegel, M., Anderson, J., Aparicio, A., King, I., Majewski, S., Mar\' \i n-Franch, A., Piotto, G., Reid, I. N., \& Rosenberg, A. 2007, \aj, in press
\bibitem[Serenelli \& Weiss(2005)]{ser}Serenelli, A. \& Weiss, A. 2005, \aap, 442, 1041
\bibitem[Sirianni et al.(2005)]{sir}Sirianni, M. et al. 2005, \pasp, 117, 1049
\bibitem[Spergel et al.(2003)]{sper}Spergel, D. N., et al. 2003, \apjs, 148, 175
\bibitem[Sweigart(1973)]{swei}Sweigart, A. 1973, \aap, 24, 459
\bibitem[Thoul et al.(1994)]{tho}Thoul, A. A., Bahcall, J. N., \& Loeb, A. 1994, \apj, 421, 828
\bibitem[Vandenberg et al.(2006)]{vdb2}Vandenberg, D. A., Bergbusch, P. A., \& Dowler, P. D. 2006, \apjs, 162, 375
\bibitem[Vandenberg \& Clem(2003)]{vdb}Vandenberg, D. A. \& Clem, J. L. 2003, \aj, 126, 778
\bibitem[Vandenberg et al.(2000)]{vdb3}Vandenberg, D. A., Swenson, F. J., Rogers, F. J., Iglesias, C. A., \& Alexander, D. R. 2000, \apj, 532, 430
\bibitem[Yi et al.(2004)]{yi}Yi, S. K., Demarque, P., \& Kim, Y.-C. 2004, \apss, 291, 261
\bibitem[Yi et al.(2003)]{yi2}Yi, S. K., Kim, Y.-C., Demarque, P. 2003, \apjs, 144, 259
\end{thebibliography}
\end{document}